    \documentclass[
    preprint,
    nofootinbib,
    amsmath,amssymb,
    aps,
    pra,
    ]{revtex4-1}
    \usepackage{graphicx}
    \usepackage{dcolumn}
    \usepackage{bm}
    \usepackage{hyperref}
     \usepackage{mathtools}
\usepackage[usenames, dvipsnames]{color}
\usepackage{setspace}
\DeclarePairedDelimiter\ket{\lvert}{\rangle}
\DeclarePairedDelimiterX\braket[2]{\langle}{\rangle}{#1 \delimsize\vert #2}

\newcommand{\nocontentsline}[3]{}
\newcommand{\tocless}[2]{\bgroup\let\addcontentsline=\nocontentsline#1{#2}\egroup}

\begin{document}
    	\title{Interacting Qubit-Photon Bound States with Superconducting Circuits}
    	
    	\author{Neereja M. Sundaresan}
    	\affiliation{%
    		Department of Electrical Engineering, Princeton University, Princeton, New Jersey 08544, USA
    	}%
    	\author{Rex Lundgren}%
    	\affiliation{%
    	Joint Quantum Institute, NIST/University of Maryland, College Park, Maryland 20742, USA
    	}%
	\author{Guanyu Zhu}%
    	\affiliation{%
    	Joint Quantum Institute,
    	NIST/University of Maryland, College Park, Maryland 20742, USA
    	}%
    	\author{Alexey V. Gorshkov}
    	\affiliation{%
    	Joint Quantum Institute and Joint Center for Quantum Information and Computer Science,
    	NIST/University of Maryland, College Park, Maryland 20742, USA
    	}%
    	\author{Andrew A. Houck} 
    	\thanks{Corresponding author: aahouck@princeton.edu}%
    	\affiliation{%
    		Department of Electrical Engineering, Princeton University, Princeton, New Jersey 08544, USA
    	}%
    	\date{\today}

    	\begin{abstract}	
Qubits strongly coupled to a photonic crystal give rise to many exotic physical scenarios, beginning with single and multi-excitation qubit-photon dressed bound states comprising induced spatially localized photonic modes, centered around the qubits, and the qubits themselves. The localization of these states changes with qubit detuning from the band-edge, offering an avenue of in situ control of bound state interaction. Here, we present experimental results from a device with two qubits coupled to a superconducting microwave photonic crystal and realize tunable on-site and inter-bound state interactions. We observe a fourth-order two photon virtual process between bound states indicating strong coupling between the photonic crystal and qubits. Due to their localization-dependent interaction, these states offer the ability to create one-dimensional chains of bound states with tunable and potentially long-range interactions that preserve the qubits' spatial organization, a key criterion for realization of certain quantum many-body models. The widely tunable, strong and robust interactions demonstrated with this system are promising benchmarks towards realizing larger, more complex systems of bound states.
	\end{abstract}
    	
    	\maketitle
    	
	
	In the strong-coupling domain, a qubit coupled to a photonic band edge forms an exponentially localized photonic mode at the qubit position, which together with the qubit forms a qubit-photon dressed bound state \cite{John1991, John1991a,John1994,John1995, Calajo2016m}. Photonic crystals are natural avenues to realize these bound states due to their intrinsically tailorable band structure, and characteristic Bloch mode electric field distribution \cite{Joannopoulos2008m} which enables access to strong coupling with qubits \cite{Goban2014, Goban2015, Gonzalez-Tudela2015, Hood2016m, Liu2016m}. Bound states in multi-qubit photonic crystal devices are an ideal platform to study many-body quantum optics in one-dimensional systems \cite{Munro2016, Calajo2016m, Shi2016,Hung2016,Douglas2015, Sanchez-Burillo2017, Asenjo-Garcia2017a}. Unlike many qubits coupled to a common cavity mode but similar to the case of some optical multimode cavities \cite{Kollar2017,Vaidya2017}, coupling to a band edge creates bound states that intrinsically preserve the spatial organization of qubits, offering the ability to create one-dimensional chains of bound states with tunable and potentially long-range interactions. The promise of engineering interaction profiles beyond the intrinsic flip-flop with additional microwave drive tones further opens the possibility of simulating a wide range of quantum spin models in future devices \cite{Douglas2015}. In this paper, we demonstrate and characterize the underlying, fundamental tunable on-site and inter-bound state interactions in a two-qubit, superconducting microwave photonic crystal device.
		
\begin{figure}
\centering
\includegraphics[scale=1.45]{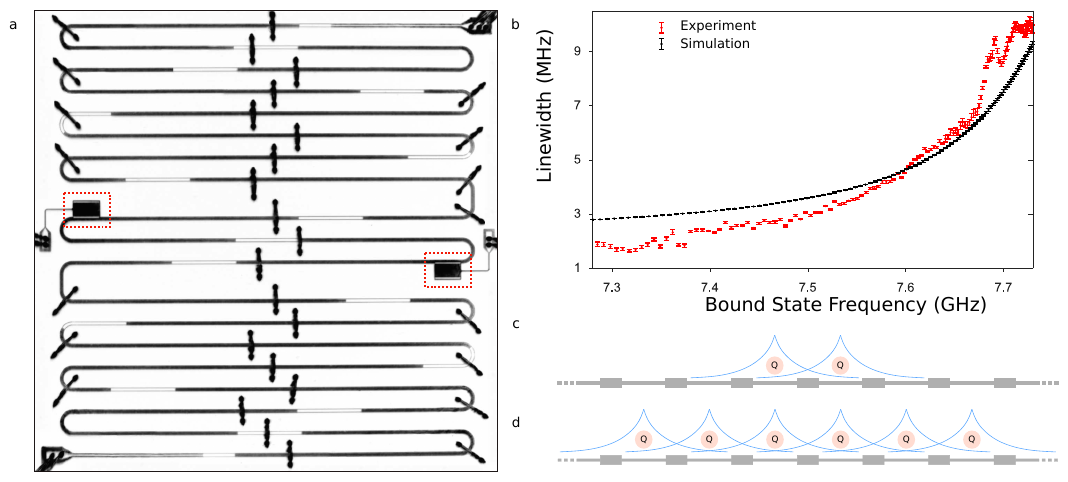}
\caption{\textbf{A platform for interacting dressed bound states} \textbf{a} A 16-site microwave photonic crystal is realized by alternating sections of high and low impedance coplanar waveguide. Two transmon qubits (multi-level, anharmonic energy ladder) are in neighboring unit cells in the middle of the device, centered in the high impedance sections for maximal coupling to the band edge at 7.8 GHz (all values presented in units of $(2\pi)$Hz. i.e. - $\omega_{BE} = 7.8 ~(2\pi)$GHz). For this experiment, the pass band (band gap) refers to states above (below) the band-edge frequency. Each qubit is individually tunable in frequency via a local flux bias line. \textbf{b} Bound state linewidth, an indirect measure of localization, varies with qubit frequency. The wide range over which photon localization can be tuned indicates the feasibility of realizing a chain of strongly interacting bound states. Experimentally measured and simulated linewidths are shown in red and black, respectively. \textbf{c} The interaction between bound states will be determined by overlap of their localized photonic envelopes with the qubits. \textbf{d} In a larger system, this localization-length-dependent interaction of the bound states would preserve the importance of the spatial organization of qubits in determining the many-body structure of the interactions.}
 \label{Fig1}
\end{figure}

	A single dressed bound state, seeded by a single qubit in a crystal, is itself a unique avenue of study. Liu et al. first directly detected such a bound state in a stepped-impedance microwave crystal coupled to a single transmon qubit \cite{Liu2016m}. That work characterized the dependence of localization length on detuning between the qubit and the band edge and further confirmed the existence of the localized state in the bandgap when the bare qubit is in the passband - an unmistakable signature of non-Markovianity (see Supplement). State localization is tunable in situ with frequency through a range determined by device parameters, including qubit-waveguide coupling and band curvature. Compared with previous work, we attain increased localization in this device  (Fig.\ \ref{Fig1}b) due mainly to a flatter band dispersion, realized by tailoring the unit cell of the photonic crystal (see Supplement). The bound state localization length in this device is still widely tunable, which is critical for realizing strong, tunable interaction between spatially separated bound states. As the different coupling regimes translate to dramatically altered system behavior \cite{Calajo2016m}, it is important to determine which domain our system falls under. In systems such as the one presented here, qubit emission into the waveguide being larger than the other decay rates (coherent atom-photon interaction rates larger than decay rates) is the minimal coupling criterion, upon which the dressed bound state within the gap can be spectrally resolved \cite{Calajo2016m}. The strong coupling criterion corresponds to the situation where a bare qubit resonant with the band edge gives rise to a bound state that is shifted from the band edge by more than the bound state's linewidth \cite{Calajo2016m, Liu2016m}. In our finite system, we observe $\sim 250$ MHz separation between the bound state and the band edge with bound state linewidth of $4$ MHz when a qubit is resonant with the band edge, thus firmly reaching the strong coupling condition (see Fig.\ \ref{Fig1}b and Supplemental Fig.\ S4a). By fabricating two qubits in the photonic crystal (Fig.\ \ref{Fig1}a), we realize multiple, spectrally-resolvable bound states and can study inter-bound state interaction. 
	
	The nature of inter-bound state interaction makes this platform intrinsically well-suited for investigating one-dimensional chains of bound states (see Fig.\ \ref{Fig1}c). Realizing sizable chains is possible by increasing the number of unit cells - a property that does not impact the Bloch mode distribution or band dispersion. Thus qubits can be in separate unit cells but realize near identical coupling to the band edge. As the strength of inter-bound state interaction depends on the spatial overlap of the photonic wave-functions with the qubits, the distance separating qubits (set by device design) is directly mapped into the interactions of the system, maintaining the chain-like interaction pattern. 

	Controlling photon-mediated interaction between superconducting qubits has been demonstrated in other one-dimensional systems - such as two qubits in a cavity \cite{Majer2007a, DiCarlo2009} or in a linear waveguide \cite{VanLoo2013}. However, in these cases the distance between the qubits was effectively eliminated (i.e. standing wave interaction in a cavity) or otherwise reduced (modulo wavelength in a linear dispersion waveguide). Thus photonic crystals and tunable bound states offer a fundamentally distinct form of interaction.  
	
\begin{figure}
\centering
\includegraphics[scale=1]{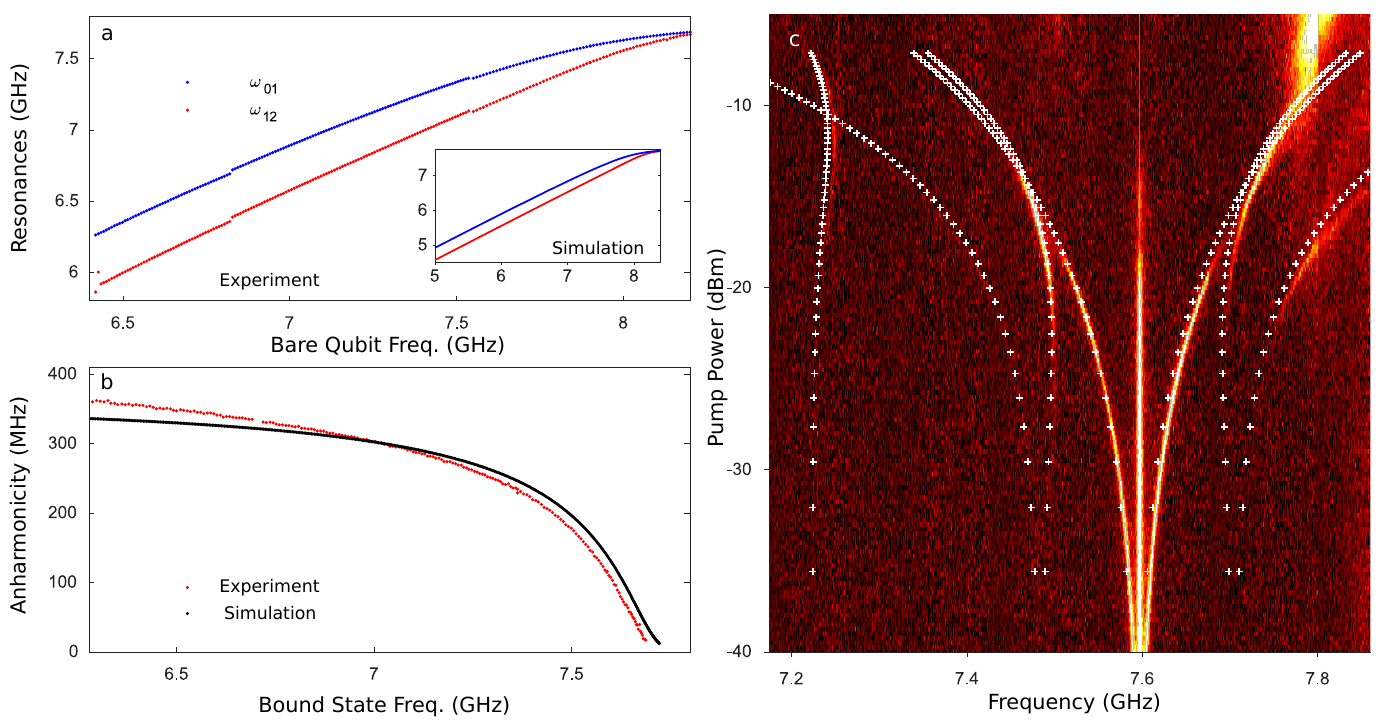}
\caption{\textbf{Probing the bound state energy levels} \textbf{a} The anharmonicity of the bound state is dependent on bare-qubit frequency, demonstrating a tunable on-site interaction strength. In blue (red), the first (second) transition of the bound state is measured across a range of bare qubit frequencies (inset - simulation).  \textbf{b} Decreasing anharmonicity with increasing bound state frequency shown in red for experimental data and black for simulation. \textbf{c} Emission spectrum of a resonantly driven ($\sim$ 7.59 GHz) bound state (induced by a qubit at 7.9 GHz, which is above the band edge located at 7.8 GHz) as a function of drive power. At low drive power, only the Mollow triplet is observed. With increasing power we see four additional sidebands, two on either side of the original Rabi sidebands, which together are the transitions between the three lowest levels of the anharmonic bound state ($\ket{0}$, $\ket{1}$, $\ket{2}$). The white crosses are from numerical simulations (see Supplement). We have included five qubit levels in our simulation. See text for the discussion of the seventh sideband around 7.25 GHz.
}
\label{Fig2}
\end{figure}

	In addition to determining localization length, the frequency of the bound state also determines on-site interaction strength. In Figs.\ \ref{Fig2}a and \ref{Fig2}b, we characterize the dependence of the transition frequencies between the three lowest levels of the bound state on bare qubit frequency, and observe the steady reduction in bound state anharmonicity from over 350 MHz to 0 MHz as the qubit is tuned from deep in the bandgap to the passband. This is dramatically more than the $\sim 10\%$ modification of qubit anharmonicity with frequency expected when a qubit is strongly coupled to a cavity mode \cite{Nigg2012}. 
	
      Therefore, while we may treat the one-excitation and two-excitation bound states as first ($\ket{1}$) and second ($\ket{2}$) excited states of a new effective anharmonic qubit \cite{Liu2016m}, it is important to note that this effective qubit differs in frequency and anharmonicity from the bare (multi-level transmon) qubit. Defining the three lowest bare qubit levels as $\ket{0}_q$, $\ket{1}_q$, and $\ket{2}_q$, here the two-excitation bound state is largely due to the coupling of the second qubit transition ($\ket{1}_q\leftrightarrow\ket{2}_q$) with the band edge rather than other multi-photon effects  \cite{Shi2016} (see Supplement). 
      
       Numerical simulations, modeling the photonic crystal as a coupled cavity array with free parameters fit to match the band curvature from the dispersion relation \cite{Biondi2014am, Calajo2016m}, show similar dependence of anharmonicity on detuning (see Fig.\ \ref{Fig2}a inset and Fig.\ \ref{Fig2}b). Unlike the transfer matrix method \cite{Pozar2005m, Shen2006m,Shen2009}, this approach can extend beyond the single-excitation manifold to capture the higher levels of the bound state, as well as the Lamb shift of the qubit frequency, observed when including next-nearest neighbor hopping between coupled cavities. Each qubit is modeled as a three-level ladder with negative anharmonicity, and with the $\ket{0}_q \leftrightarrow \ket{1}_q$ and $\ket{1}_q\leftrightarrow\ket{2}_q$ transitions coupled with amplitudes $g$ and $g\sqrt{2}$, respectively, to its local cavity-site. It is critical to include level $\ket{2}_q$ to accurately reproduce the two-excitation manifold observed in experiment.
	
The tunable level structure also emerges in the emission spectrum of a continuously driven bound state (Fig.\ \ref{Fig2}c), induced by a single qubit with bare frequency above the band edge. At low drive amplitude or Rabi frequency, transmission across the crystal via the bound state exhibits antibunching \cite{Hafezi2012}, consistent with single photon transport of a two-level system and resonant pump (see supplement) \cite{Astafiev2010, Abdumalikov2011, Hoi2012m}. When the Rabi frequency is on the same order as the anharmonicity, the bound state can no longer be approximated as a two-level system. 
In this domain, the steady state will be a mixture of the three eigenstates obtained by diagonalizing the drive Hamiltonian in the Hilbert space spanned by $\ket{0}$,$\ket{1}$, and $\ket{2}$. Transitions between all three eigenstates result in six sidebands \cite{Koshino2013}. These six sidebands are visible in Fig.\ \ref{Fig2}c, though emission intensity varies greatly among them due to eigenstate population. 
A seventh transition is evident in the data (7.25 GHz in Fig.\ \ref{Fig2}c). This additional transition is due to the fourth effective qubit level ($\ket{3}$) while its curvature is reproduced by including a fifth effective qubit level ($\ket{4}$)  in our numerical simulations (see Supplement). Crucial to reproducing this transition in our theoretical simulations is taking into account that the bound state level structure cannot be defined by a single anharmonicity, i.e. given the anharmonicity, $\Delta=\omega_{12}-\omega_{01}$, of the bound state, the frequency of the fourth level of the bound state is not simply given by $3\omega_{01}-3\Delta$ as is expected for a transmon \cite{Koch2007}.


\begin{figure}
\centering
\includegraphics[scale=1.1]{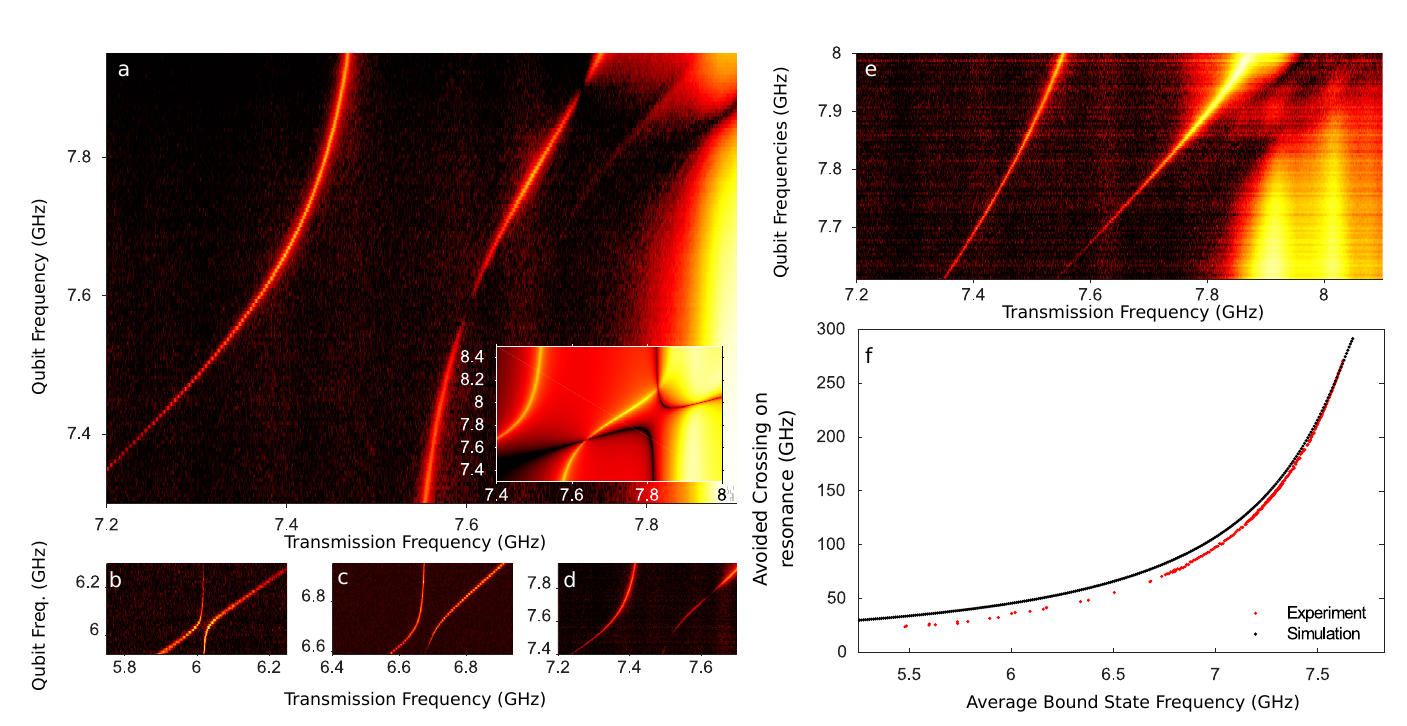}
\caption{\textbf{Interacting Bound States -} Interaction between bound states is characterized by the avoided crossing seen in transmission while tuning one qubit (y-axis) through resonance with the other (fixed). \textbf{a} An avoided crossing of 240 MHz is observed when the fixed qubit is at 7.73 GHz. The two points where transmission amplitude of a bound state dims are understood as the bound state peak being resonant with the qubit frequency. \textbf{a (inset)} Hopping model simulation of the one-excitation manifold is consistent with experimental observation. The lamb shift in the hopping model originates from next-nearest neighbor interaction between coupled cavities. \textbf{b, c, d} Tunable bound state interaction strength is illustrated in example bound state avoided level crossings for a fixed qubit whose bare frequency is circa 6.125, 6.75, and 7.625 GHz. As qubits are detuned further from the band edge, bound states are more tightly localized, reducing overlap and thus reducing interaction. \textbf{e} Measuring bound state separation as qubits are simultaneously tuned, maintaining resonance, shows changing bound state interaction strength with frequency. \textbf{f} Bound state avoided crossing as a function of average bound state frequency. A steady reduction in interaction strength occurs with increasing detuning from the band edge (moving deeper into the bandgap) due to increasing localization of the bound states. Hopping model simulation (black) captures this detuning-dependent behavior. 
}
\label{Fig3}
\end{figure}

    
    We observe the flip-flop interaction between the two spatially separated bound states by measuring the avoided crossing in transmission when the bound states are tuned into resonance. As these qubits are a fixed distance apart (9 mm) and there is negligible direct capacitive coupling, the strength of the flip-flop interaction will be entirely determined by the overlap of the localized photonic mode of one qubit with the other qubit, controllable here via the qubit frequencies. 
    
    
        In Fig.\ \ref{Fig3}a, the frequency of one qubit is held constant while the other is tuned through resonance. Measuring transmission at the single photon level reveals an avoided crossing between the $\ket{01}$ and $\ket{10}$ levels of the coupled dressed bound states. Transmission amplitude of a bound state dims when the bound state and bare qubit are near resonance (see Fig.\ \ref{Fig3}a inset). From this plot, we can extract a resonant bound state - bound state interaction of 120 MHz for a 7.73 GHz bare qubit frequency. In comparison, Fig.\ \ref{Fig3}b-d shows reduced interaction strength when both qubits are further detuned from the band edge, 6.125, 6.75, and 7.625 GHz, respectively, for the fixed qubit frequency. 
        
         To characterize this aspect of the two bound state interaction, we map the magnitude of the avoided crossing as a function of detuning. In Fig.\ \ref{Fig3}e and \ref{Fig3}f, the qubits are maintained on resonance with one another while being simultaneously tuned through the bandgap. Theoretical modeling (Fig.\ \ref{Fig3}f) shows experimental data to be consistent with localized photonic states and with interaction via wavefunction overlap. In the limit where the qubit is deep in the gap, the Markovian approximation holds. Here the localized mode and flip-flop interaction both have the same distance dependence $e^{-x/L}$ where $L=a\sqrt{\frac{\alpha}{\delta}}$ is the localization length,  $\delta$ is the detuning between the bare qubit and the band edge, $a$ is the unit cell size, and the band-edge dispersion is $\omega_k = \omega_0+\alpha a^2{(k - k_0)^2}$ (see Supplement). The corresponding flip-flop interaction Hamiltonian is $H \propto \sum_{j,l} S_i^+S_j^- (-1)^{|x_i-x_j|/a}e^{-|x_i-x_j|/L}$.

\begin{figure}
\centering
\includegraphics[scale=1]{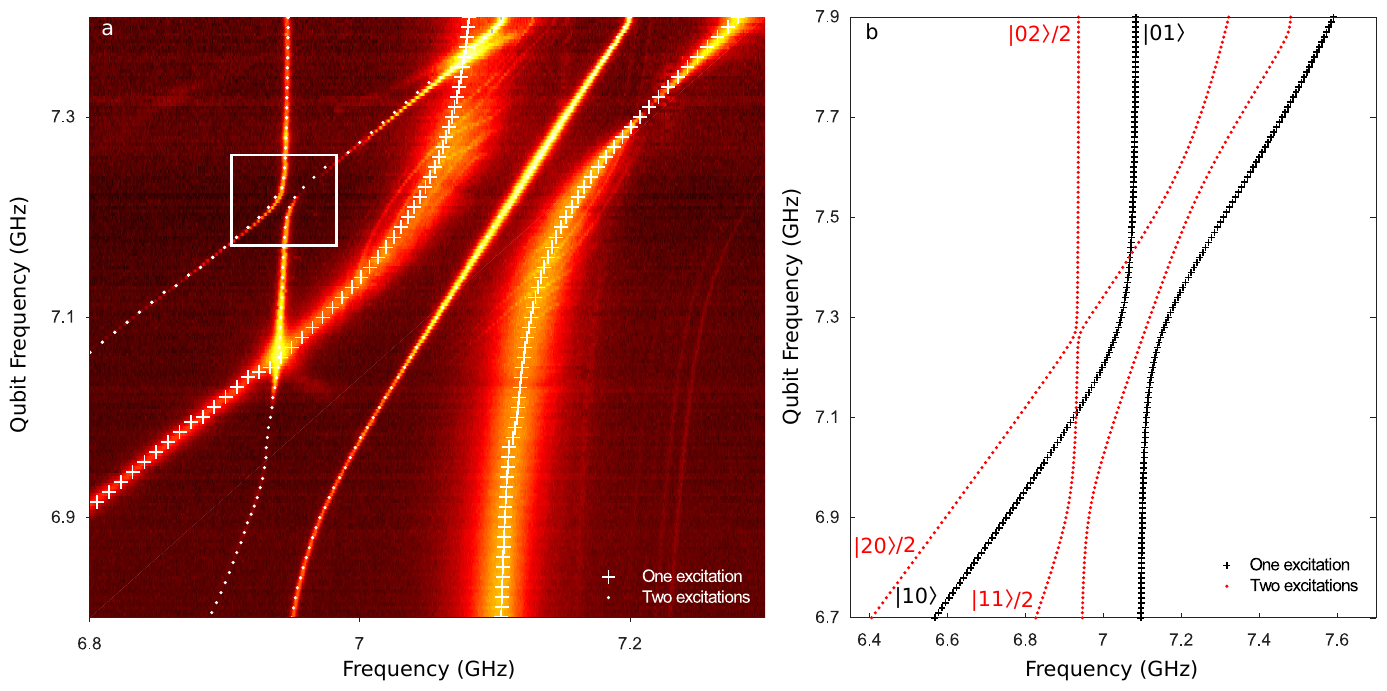}
\caption{\textbf{Interaction between two-excitation levels of two bound states} \textbf{a} Spectroscopic measurement while tuning one bound state through the other (qubit fixed at 7.2 GHz), reveals survival of strong interaction into the two-excitation manifold. Crossed and dotted lines are guides to the eye to discern the levels belonging to the first ($\ket{01}$ and $\ket{10}$) and second ($\ket{02}$, $\ket{20}$, and $\ket{11}$) excitation manifolds, respectively. In addition to the  $\ket{02}$($\ket{20}$) and $\ket{11}$ avoided level crossing, we also detect a two photon virtual interaction between $\ket{02}$ and $\ket{20}$ (white box). This interaction - fourth order in coupling $g$ - manifests itself in avoided level crossings up to and exceeding 20 MHz. For comparison, the $\ket{02}$($\ket{20}$) - $\ket{11}$ and  $\ket{01}$ - $\ket{10}$ interactions are second order in $g$ and thus both are significantly stronger. \textbf{b} Numerical simulation for fixed bare qubit frequency of 7.27 GHz, with the one (two) excitation manifold in black (red).} 
\label{Fig4}
\end{figure}

To further study tunable on-site interaction, we probe the interacting bound states beyond the one-excitation manifold using spectroscopic measurements (see Fig.\ \ref{Fig4}a). Similar to spectroscopy of qubits in cavities, we can use transmission at the band edge to help detect bound state transitions, a technique that provides sharper contrast compared to transmission measurement for the more highly localized bound states and allows detection of higher dressed transitions, such as the transition between $\ket{0}$ and $\ket{2}$ driven by two photons of frequency $\omega_{02}/2$. 
        
        With this technique we detect interaction between $\ket{02}$, $\ket{20}$, and $\ket{11}$ of the coupled bound states, observed as avoided level crossings. In addition to the single-photon exchange interaction between $\ket{02}$ ($\ket{20}$) and $\ket{11}$ \cite{DiCarlo2009}, remarkably we measure the two-photon virtual interaction between $\ket{20}$ and $\ket{02}$, despite the fact that this process is 4th order in coupling $g$ (see Supplement). This two-photon interaction shows consistent dependence on detuning: increasing in strength (from 0 to over 10 MHz) as the bound states shift towards the band edge and the states become more delocalized. Numerical simulations (Fig.\ \ref{Fig4}b) are consistent with experimental data and capture the relative magnitudes of interaction between levels as well as frequency dependence on coupling strengths. Observation of this small interaction highlights the overall strength of inter-bound state coupling possible via overlap alone.   
        
The widely tunable on-site and inter-bound state interactions demonstrated with this device and consistent theoretical simulations are promising benchmarks towards realizing larger, more complex systems of bound states that can mediate both local and long-range interactions. Beyond stepped impedance coplanar waveguides, there are undoubtably numerous ways to realize superconducting microwave photonic crystals, including lumped element or Josephson junction based designs, that are equally compatible with superconducting qubits. Regardless of the platform, behavior of bound states due to qubit-band edge coupling will mirror the behavior characterized in this work - elevating this platform above any single experimental design choice.  

While the bound states were centered in neighboring unit cells in this device, this is not a limitation or requirement for future experiments as the range of localization can be accordingly set via the basic crystal parameters, as seen by comparing bound state linewidths measured here with those reported previously \cite{Liu2016m}. Therefore, one can realize a one-dimensional chain of bound states in a moderately-sized photonic crystal, where individual control over the qubits would allow dialing up or down long-distance interaction between sets of qubits. 

In this work the interactions were in situ tunable via qubit frequency (DC flux), a static quantity on the timescale of the bound state lifetime. Dynamically controllable interactions would introduce an additional tool for designing and manipulating spin Hamiltonians \cite{Douglas2015}. One method for realizing this type of fast timescale control is flux-pumping, a technique involving microwave frequency modulation of the qubit frequency along the flux bias line \cite{Beaudoin2012,Strand2013,McKay2016,Naik2017}. Another potential pathway would use an auxiliary microwave field through the crystal itself. Here the qubits could be maintained on resonance deep in the bandgap such that there is minimal interaction via bound state overlap. A single RF control tone can be turned on to induce a transition close to the passband, thus re-dressing the bound states into new, effective bound states with interaction strength depending on properties of the microwave drive. The addition of several drives or precisely shaped microwave pulses (made possible by commercial high-speed arbitrary microwave waveform generators \cite{Raftery2017}) promise not only changing interaction strength but also modifying the shape of the interaction itself - from an exponential to a sum of exponentials - leading to a wide range of possibilities including power-law-decaying interactions \cite{Douglas2015}. These supplementary forms of tunable control would expand the ability of the qubit-photonic crystal platform to realize a broader class of tunable spin models. 

\vspace{1mm}
   \textbf{Acknowledgements}
    	We thank D. Chang, J. Douglas, A. González-Tudela, P. Rabl, and Y. Wang for valuable theoretical conversations, M. Fitzpatrick, Y. Liu, A. Vrajitorea, G. Zhang  for insightful experimental discussions, and C. Eichler, A. Gyenis, A. Kollar, Z. Leng for helpful conversations and technical support on setting up the correlation measurement.
	N.M.S and A.A.H acknowledge support from NSF (PHY-1607160),  ARO MURI (W911NF-15-1-0397), Princeton University, and NDSEG. R.L. and A.V.G. acknowledge support by NSF QIS (PHY-1415616), ARL CDQI (W911NF1520067), NSF PFC at JQI (PHY-1430094), AFOSR (FA95501510173), ARO (W911NF1410599), and ARO MURI (W911NF1610349). G. Z acknowledges support from ARO-MURI (023947-002) and YIP-ONR PFC. There are no competing financial interests.
\vspace{1mm}


   \vspace{2mm}
 
\newpage

\newcommand{\beginsupplement}{%
        \setcounter{table}{0}
        \renewcommand{\thetable}{S\arabic{table}}%
        \setcounter{figure}{0}
        \renewcommand{\thefigure}{S\arabic{figure}}%
        \setcounter{equation}{0}
        \renewcommand{\theequation}{S\arabic{equation}}
     }

\vspace{2mm}
\textbf{{Supplemental Information: Interacting Qubit-Photon Bound States with Superconducting Circuits}}

\ifpdf
\DeclareGraphicsExtensions{.pdf, .jpg, .tif}
\else
\DeclareGraphicsExtensions{.eps, .jpg}
\fi



    	\beginsupplement


   \section{Motivation for using Photonic Crystals} \label{expDetails}

To realize a dressed bound state between a qubit and a band edge, we must couple a qubit to a photonic band edge, where a band edge consists of photonic states corresponding to the transition between a stopband and a passband that are  "slow-light" due to  reduced group velocity.  Bandgaps and passbands are not unique to photonic crystals - we come across them in numerous other structures such as waveguides (near cutoff frequency) or aperiodic filters.

To motivate the benefit of photonic crystals, let us consider using an aperiodic filter instead. Filters are ubiquitous in the microwave domain with many established design methods that trade off optimizing various parameters such as roll-off or passband ripple. Stepped impedance filters are a standard model for implementing filters, where each impedance step is chosen precisely to meet filter design constraints with no periodicity requirement \cite{Pozar2005}. Due to this, aperiodic filters may also be more sensitive to fabrication errors than periodic filters.

The next requirement is to couple a qubit. However the lack of periodicity transfers also to the electric field distribution (no Bloch modes), and so we must numerically calculate the optimal location to place the qubit and recalculate for every modification of filter design. While this seems feasible for a single qubit, the lack of Bloch modes is highly problematic from a scalability standpoint as it is not guaranteed that we can couple multiple qubits (or even just two), with near identical coupling strengths, to that band edge. 

Furthermore, from a theoretical perspective, periodic crystals are simpler, cleaner structures that are described in the infinite limit by dispersion relations in momentum space, providing useful insight for predicting system behavior. Therefore, while it may be possible to realize dressed bound states with a qubit in an aperiodic structure, the benefits from using a periodic structure far outweigh the likely larger device footprint. 

	A photonic crystal is an electromagnetic structure formed by periodic modulation of the dielectric constant. This results in dispersion relations characterized by energy bands - alternating bandgaps ("forbidden energies") and passbands (continuous density of states). The electric field distribution is characterized by Bloch modes, allowing for the position of qubits at locations that optimize coupling to the band edge \cite{Joannopoulos2008}. Engineering band edges via finite-size photonic crystals has been demonstrated in many systems, including nanophotonic structures \cite{Lund-Hansen2008,Yu2014} and superconducting microwave coplanar waveguides \cite{Bronn2015, Liu2016}, and tremendous progress towards integration with ultracold atoms and superconducting qubits, respectively, has been reported \cite{Liu2016,Hood2016}. There are other ways to create superconducting microwave crystals, including using Josephson junction arrays or lumped element circuits.

	Our approach to creating an effective 1D microwave photonic crystal is periodically alternating sections of high and low impedance coplanar waveguides (CPWs). With CPWs, the impedance can be easily changed via the center-pin width and the center-pin to ground plane (the gap) distance. 
	
	We define a unit cell as a high impedance section of length $L_{hi}$ and impedance $Z_{hi}$ sandwiched between two sections of low impedance of lengths $L_{lo}/2$ and impedances $Z_{lo}$ (for symmetry purposes). With a periodic modulation, there are naturally many gaps in the band structure. We chose to more strongly couple the qubit to the second band edge, rather than the first, because it has a smoother passband while still having a steep roll-off. 

\subsection{Crystal Simulation and Parameters for Implementation} \label{dispRel}
We can use the unit cell to calculate the band structure or dispersion relation for an infinite crystal. While we can never make an infinite crystal, calculating the dispersion relation is a very useful starting point and gives us insight into crystal parameters such as band curvature. To a good approximation, the phase velocities in the high and low impedance CPW sections are effectively equal ($v_{p, high}\sim v_{p, low} \sim v_{p} \approx 1.248\times 10^{8}$ m/s). This yields

\begin{equation}
\begin{aligned}
\cos(\frac{\omega_k L_{lo}}{v_p})\cos(\frac{\omega_k L_{hi}}{v_p}) - \frac{1}{2}(\frac{Z_{hi}}{Z_{lo}} + \frac{Z_{lo}}{Z_{hi}})\sin(\frac{\omega_k L_{lo}}{v_p})\sin(\frac{\omega_k L_{hi}}{v_p}) = \cos(k(L_{lo}+L_{hi})),
\end{aligned}
\end{equation}
where $k$ is the momentum and $\omega_k$ is the dispersion. 

We determine the band structure dependence on $L_{lo}$ and $L_{hi}$ for $Z_{lo} \sim 25 \Omega$ and  $Z_{hi} \sim 125 \Omega$. We look to optimizing the trade-offs across four parameters: the frequency of the band edge, the width of the bandgap, the width of the passband, and the curvature of the band. From these simulations, we see that small changes in unit cell impedances do not lead to significant changes in the band dispersion. 

\begin{figure}
\centering
\includegraphics[scale=1.1]{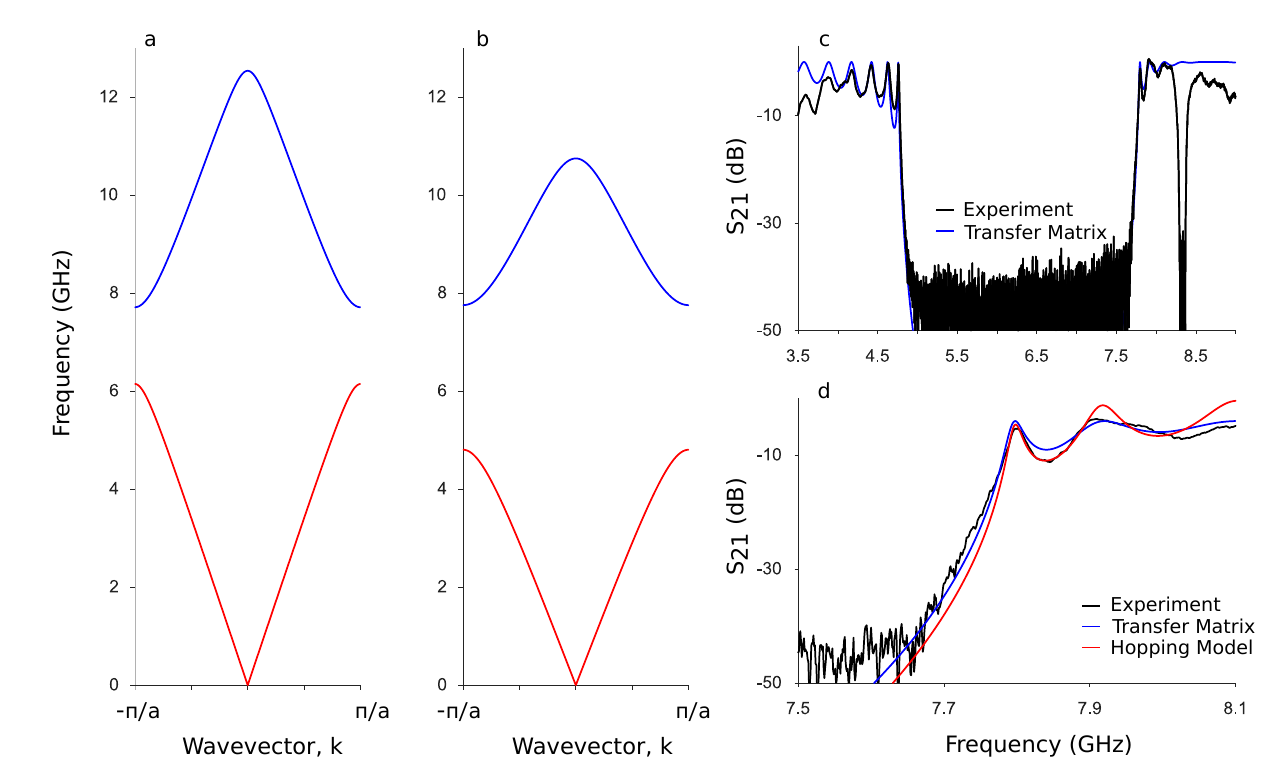}
\caption{First two bands of simulated dispersion for  $Z_{lo} = 25 \Omega$,  $Z_{hi} = 124 \Omega $  and \textbf{a} $L_{lo} = 0.45$ mm,  $L_{hi} = 8$ mm (parameters from \cite{Liu2016}), and \textbf{b}  $L_{lo} = 1.2$ mm,  $L_{hi} = 7.8$ mm (parameters from present paper). \textbf{c} Overlay of simulated $S_{21}$ from the transfer matrix method (blue; same parameters as \textbf{b}) and measured high-power $S_{21}$ (black) shows good agreement in bare crystal characteristics. \textbf{d} Overlay of simulated $S_{21}$ from the transfer matrix method (blue; same parameters as \textbf{b}) and from the hopping model (red; with $\kappa=1$ GHz and $\kappa_0=4$ MHz) and measured high-power $S_{21}$ (black) near the band edge. Drop in transmission circa 8.3 GHz is due to TWPA dispersion, not device defect. 
 }
\label{crystal}
\end{figure}

For comparison, Figs.~\ref{crystal}a and b show the simulated dispersion for unit cell parameters in Liu et al. \cite{Liu2016} and the present paper, respectively. The unit cell for the present paper was chosen to have a flatter band dispersion (analogous to effective mass), $\alpha$, so as to realize more localized bound states.

	While the dispersion relation assumes an infinite system, crystals of small, finite length have been shown to realize well-resolved gaps in dispersion where transmission is suppressed and bands where transmission is unimpeded. From an experimental perspective, we use transmission-based measurements to probe the states. In the investigation of bound states, the finite size of the crystal is a practical advantage: the overlaps of bound states with ends of the crystal lead to only quasi-bound states, allowing for detection through transmission measurements. Direct detection of such a state in this way was first demonstrated in Liu et al. \cite{Liu2016}.
	
	Transfer matrices are a convenient and accurate method for bare crystal simulation that incorporates both the exact number of cells and boundary conditions \cite{Shen2006}. A convenient metric for comparison is transmission across the device, $S_{21}$. In Figs.~\ref{crystal}c and d, comparison of the measured $S_{21}$ of the bare crystal (taken at powers high enough to saturate qubit effects) and the calculated $S_{21}$ from transfer matrices shows agreement \cite{Liu2016}. 
	
\subsection{Realistic Parameters and Experimental Aside} \label{params}

	The device is fabricated on a $\sim 430 \mu$ m thick c-plane sapphire substrate, on which we sputter approximately 200nm of niobium. The photonic crystal is patterned using a direct-write laser writer, followed by a dry $SF_6$ reactive ion etch to transfer the pattern. To ensure our waveguides are reasonably sized and can be reliably fabricated using photo-lithography, we are limited to impedances between approximately 25 $\Omega$ to 125 $\Omega$.  

For this device, one unit cell consisted of a high impedance section ($Z_{hi} = 124\Omega$, $L_{hi} = 7.8$ mm) and a low impedance section ($Z_{lo} = 25\Omega$, $L_{lo} = 1.2$ mm). Impedance estimates are obtained by fitting the measured spectrum with transfer matrix simulations; the dispersion is robust to small impedance deviations in the fabricated sample. We fit 16 unit cells on a 10 mm x 10 mm sapphire chip. While more unit cells could fit on the chip, we saw experimentally that we must wire-bond extensively on-chip (to connect ground planes) to create clean bandgaps and passbands and included this requirement into the design. By switching to air or dielectric-supported bridges in the future we would be able to shrink the device footprint or include more unit cells in the same size device. 

	To integrate with the standard measurement setup and qubit parameters, we choose unit cell lengths such that the second band edge falls between 7 GHz and 8 GHz. The frequency of the band edge was designed to be at 7.8 GHz to match with the traveling wave parametric amplifier (TWPA, dispersion feature around 8.3 GHz) to provide good amplification for frequencies in the vicinity of the band edge as well as deep in the bandgap. The width of the bandgap is from 4.75 GHz to 7.8 GHz and needed to be large enough such that we have a wide range to tune the qubit frequency over, which would in turn result in a wide range of accessible localization lengths.  The width of the (second) passband needs to be sufficiently large such that we can ignore the third band edge and we can make the approximation that the curvature of the band (near the band edge) is quadratic. Finally as the curvature of the band plays a role in determining the localization length, to make the states more localized, we chose a shallower band curvature (compare red plots in Fig S1a and S1b). 
		
	In future iterations, one may consider altering boundary conditions specific to the desired application. For example, methods for impedance matching such as tapers or quarter-wave transformers \cite{Pozar2005} are straightforward additions that will modify impedance matching at specific frequencies (such as at a band edge). These options were not pursued in this work as we wanted to study bound state properties across a range of frequencies. If one were interested in only specific frequencies or ranges, then this is a promising improvement. The device is symmetric, so one end is chosen arbitrarily to serve as the input.  As detection of the bound state is due to scattering, one may consider modifying the symmetry of the device or detecting signal from both ends of the crystal to improve collection efficiency. 

	We resorted to two-tone spectroscopy over direct transmission measurement to detect the bound state when the qubit was deep in the bandgap due to strong localization and poor signal-to-noise ratio (SNR). Choosing a less shallow band curvature will make it possible to see the bound state all the way through the gap (see Liu et al.\ \cite{Liu2016}); however, this is a trade-off as the bound-state linewidths will increase accordingly. Thus, choosing curvature and crystal parameters such that the linewidth remains as narrow as possible without internal Q or loss effects is essential. 

\section{Adding in Qubits }
In this device, we capacitively couple a transmon qubit to each of the two central unit cells of the 16-cell crystal. A transmon's anharmonic level structure is due to the nonlinear inductance of Josephson junctions and allows for selective addressing of energy level transitions. However, it is important to emphasize that our realization of a qubit does have higher energy levels set by transmon geometry, unlike the standard theoretical qubit which is synonymous with a two-level system. These higher levels are also coupled to the band edge and therefore accounting for these levels becomes important when looking beyond the first excitation sector. 

Our qubits are fully patterned with a 125kV e-beam writer, with bridge-free junctions, and are made of evaporated aluminum. The qubits were designed to have a target charging energy (from electro-magnetic simulation) of approximately 450 MHz, and emphasis was placed on maximizing coupling between the qubit and photonic crystal without significantly altering the unit cell. Finally, although identical in design, in reality the qubits will differ due to fabrication uncertainty/error. However, as coupling to the waveguide is designed to be the dominant decay channel for the qubits, bound states are robust to small variation in other parameters. 

We place qubits in adjacent unit cells (9mm separation) at the center of the crystal such that we can see significant change in interaction strength with detuning. Each qubit has a local flux bias line for independent control -  DC cross-talk is corrected for through standard calibration. These lines are low pass filtered; however, as the dominant decay of the qubit is via the crystal, this is not expected to be a limiting factor in coherence. 

To determine where to place the qubits within the unit cell such that they maximally couple to the desired band, we must calculate the electric field distribution within the unit cell, distribution determined by the Bloch modes for the crystal \cite{Joannopoulos2008}. 

For these crystal parameters, maximally coupling to the second band edge (and minimally to the first band edge) corresponds to placing the qubit in the center of long high-impedance section. Other locations within a unit cell change coupling to each band edge, which is a potentially interesting regime for future experiments. However, here we are interested in reducing the effect of the other band as much as possible. We cannot completely eliminate this coupling - experimentally we can still detect a drop in transmission in the lower passband when the qubit is resonant but this change is orders of magnitude smaller than in the second band. 
\section{Model, Transmission, and Fitting of Parameters}
In this section, we introduce the effective Hamiltonian for our system and discuss the fitting of various parameters of the Hamiltonian in detail. 
\subsection{Hamiltonian}
The Hamiltonian for the one-dimensional photonic crystal is given by $H_c=\sum_{k}\omega_ka^\dagger_k a^{\phantom{\dagger}}_k$, 
where $a^\dagger_k$ creates a bosonic excitation with momentum $k$, and $\omega_k$ is the dispersion relation of the second band of the photonic crystal, which is discussed in Sec.~\ref{dispRel}. Here, we have ignored the other bands of the photonic crystal as the qubits couple predominantly to the second band. Fourier transforming ($a^\dagger_k=\sqrt{\frac{a}{L}}\sum_{j=1}^Na^\dagger_{j}e^{ik x_j}$, where $L$ is the system size, $a$ is the unit cell size, $N$ is the number of unit cells, $x_j=aj$, and $k=\frac{2\pi}{L}n$, where $n$ is an integer) gives a hopping model with periodic boundary conditions, $H_c=\sum\limits_{i,j}J_{i,j}a^\dagger_{i} a^{\phantom{\dagger}}_{j}$, where
\begin{equation}
J_{i,j}=J_{|i-j|}=\lim_{N\rightarrow\infty}\frac{1}{N}\sum_{k}e^{ik(x_i-x_j)}\omega_k=\int_{-\pi}^{\pi}\frac{d\tilde{k}}{2\pi}e^{i\tilde{k}(x_i-x_j)/a}\omega_{\tilde{k}/a}.
\label{hopping}
\end{equation}
Here, $\tilde{k}$ is a dimensionless integration variable, and we have used the fact that $J_{i,j}=J_{j,i}$ since $\omega_k=\omega_{-k}$. In Eq.\ (\ref{hopping}) we have taken $N\rightarrow \infty$, as we only know $\omega_k$ in that limit (see Sec.\ \ref{dispRel}). 
To model our finite system, which has open boundary conditions and 16 unit cells, we use a 16 site hopping model with open boundary conditions with hopping strengths determined by Eq.~(\ref{hopping}). Using the photonic crystal parameters in Sec.~\ref{params}, we find (by numerical integration) $J_0=9.3272$ GHz, $J_1=0.7288$ GHz, $J_2=-0.0344$ GHz, $J_3=0.0178$ GHz, $J_4=-0.0034$ GHz, and $J_5=0.0014$ GHz. Unfortunately, we are unaware of an exact analytical solution for $J_{i,j}$ for our system. In our numerical simulations, we keep hopping terms up to $J_5$.  We have calculated the hopping parameters for a given set of photonic crystal parameters. A different choice of photonic cyrstal parameters would have given a different set of hopping parameters. As such, these parameters should only be understood as estimates. We briefly comment on the change in theory parameters that arises from using different photonic crystal parameters at the end of this section.

The Hamiltonian for the isolated qubits is given by 
\begin{equation}
H_{q}=\sum_{i=1,2}\sum_{n=0}^\infty\omega_{0n;i}|n\rangle_i\langle n|_i.
\end{equation}
Here, $i$ labels the qubit, $n$ labels the level of the qubit and $\omega_{0n;i}$ are the bare energy levels of the qubits. In our simulation, the number of qubit levels is truncated at five (i.e. $|0\rangle$ through $|4\rangle$) . For our experiment, $\omega_{00;i}=0$, $\omega_{02;i}=2\omega_{01;i}-\Delta$, $\omega_{03;i}=3\omega_{01;i}-3\Delta$ and $\omega_{04;i}=4\omega_{01;i}-6\Delta$, where $\Delta$ is the bare anharmonicity of the qubits, which is taken to be the same for both qubits.

We now turn to the coupling of  the qubits to the photonic crystal. To an excellent approximation, the coupling between the qubit and the photonic crystal takes place within a single unit cell. Thus, in the rotating-wave approximation, we can write the coupling term of the Hamiltonian as
\begin{equation}
H_{qc}=\sum_{i=1,2} g_ia^\dagger_{z_i}(|0\rangle_i\langle 1|_i+\sqrt{2}|1\rangle_i\langle 2|_i+\sqrt{3}|2\rangle_i\langle 3|_i+\sqrt{4}|3\rangle_i\langle 4|_i)+h.c.,
 \end{equation} 
where $z_i$ labels the position of the two qubits. In our system, the qubits are on neighboring unit cells, i.e. $z_1=\frac{N}{2}$ and $z_2=\frac{N}{2}+1$, and the coupling for each qubit, $g_i$, is different due to small experimental imperfections. The total Hamiltonian of the system is then 
\begin{equation}
H_{tot}=H_c+H_q+H_{qc}. \label{eq:htot}
\end{equation}
Finally, we note that this Hamiltonian conserves total excitation number.
 
 \subsection{Transmission Methods}
We now discuss the two theoretical methods we use to calculate transmission in the linear drive regime. Neither method relies on a weak coupling approximation between the qubits and the photonic crystal. The first method involves treating the system as an open quantum system, with loss on each site (that is site-dependent), subject to a weak drive on the first site. The largest loss terms are at the ends of the one-dimensional photonic crystal. The system can then be described by the following effective non-Hermitian Hamiltonian (in the rotating frame) with driving frequency $\omega_d$ and strength $\epsilon$,
\begin{align}
H_{eff}=\sum_{i,j=1}^{N}(J_{i,j}-\omega_d\delta_{i,j}-i\kappa_0\delta_{i,j})a^\dagger_{i }a^{\phantom{\dagger}}_{j}+ \sum_{i=1,2}(\omega_{01;i}-\omega_d-i\kappa_{q})|1\rangle_i\langle 1|_i+\nonumber \\
+\sum_{i=1,2} g_i(a^\dagger_{z_i}|0\rangle_i\langle 1|_i+a^{\phantom{\dagger}}_{z_i}|1\rangle_i\langle 0|_i)-i\kappa( a^\dagger_{1}a^{\phantom{\dagger}}_{1}+a^\dagger_{N}a^{\phantom{\dagger}}_{N})+\epsilon(a^\dagger_{1}+a^{\phantom{\dagger}}_{1}),
\end{align}
where $\kappa_q$ is the qubit halfwidth, $\kappa_0$ is a uniform contribution to photonic halfwidth, and $\kappa$ is a decay rate on the first and last sites. While there are certainly other forms of loss (such as non-uniform loss on each site), our goal is to reproduce the key features (for example, the locations of the bound state and of the transmission dip, as well as the linewidth of the bound state) using as few parameters as possible. The equations of motion for the quantum operators are
\begin{eqnarray}
\frac{\partial a^{\phantom{\dagger}}_{y}}{\partial t}=-i\bigg(\sum_{j}(J_{y,j}-\omega_d\delta_{y,j})a^{\phantom{\dagger}}_{j}+\epsilon\delta_{y,1}+\sum_{i=1,2} \delta_{y,z_i}g_i|0\rangle_i\langle 1|_i\bigg)-(\kappa_0+\kappa( \delta_{y,1}+\delta_{y,N}))a^{\phantom{\dagger}}_{y},\\
\frac{\partial (|0\rangle_i\langle 1|_i)}{\partial t}=(-i(\omega_{01;i}-\omega_d)-\kappa_q)|0\rangle_i\langle 1|_i-ig_ia^{\phantom{\dagger}}_{z_i}(|0\rangle_i\langle 0|_i-|1\rangle_i\langle 1|_i).
\end{eqnarray}
We have omitted vacuum Langevin noise from the equations of motion as this noise does not affect our calculations. Solving for the steady state of $a_{N}$ in the weak drive limit ($\langle|1\rangle_i\langle 1|_i\rangle_{ss}=0$) yields the transmission. More specifically, $S_{21}\propto \langle a_{N}\rangle_{ss}/\epsilon$.

The second method we use was introduced by Biondi et al \cite{Biondi2014a}. Here, we treat the system (which is taken to be the photonic crystal and qubits) as being connected to waveguides with linear dispersions, with velocity $v_g$, at the ends of the photonic crystal. 
In this method, we take $\kappa_q = \kappa_0 = \kappa = 0$, so that the single-excitation transmission through the system can be expressed in terms of eigenvalues and eigenvectors of the system described by $H_{tot}$ [Eq.\ (\ref{eq:htot})]  \cite{Biondi2014a}. More explicitly, the transmission for a given frequency $\omega$ is given by $|T(\omega)|^2$, where
\begin{equation}
T(w)=\frac{-2i\beta}{\Gamma_l\Gamma_r+|\beta|^2},~~~\Gamma_{l,r}=1+\frac{i}{2v_g}\sum_{n}\frac{|V_n^{l,r}|^2}{\omega-\Omega_n},~~~\beta=\frac{1}{2v_g}\sum_{n}\frac{V_n^{l}(V_n^{r})^*}{\omega-\Omega_n}.
\end{equation}
Here, $V_{n}^{l,r}=\sqrt{v_gg_w} \langle 0|a_{1,N}|n\rangle$, $\Omega_n$ and $|n\rangle$ are the eigenvalues and eigenvectors of $H_{tot}$ in the single-excitation sector, and $g_w$ is the coupling between the waveguide and the photonic crystal. 
Intuitively, transmission occurs when the single-excitation eigenstates have the probability of the photon being on both ends of the photonic crystal.

\subsection{Fitting of Parameters}
In this section, we discuss how we fit various parameters of the total Hamiltonian. The unknown parameters include $g_i$ and $\Delta$, and, for the first method, also $\kappa_0$, $\kappa_{q}$, and $\kappa$. Furthermore, $\omega_{01;i}$ is tunable but its value is unknown, and the transmission dip (visible when the bare qubit frequency is in the passband) does not, in general, occur at the bare qubit frequency unless hopping amplitudes $J_{i,j}$ beyond nearest-neighbor are negligible. 

\begin{figure}
\centering
\includegraphics[scale=1.1]{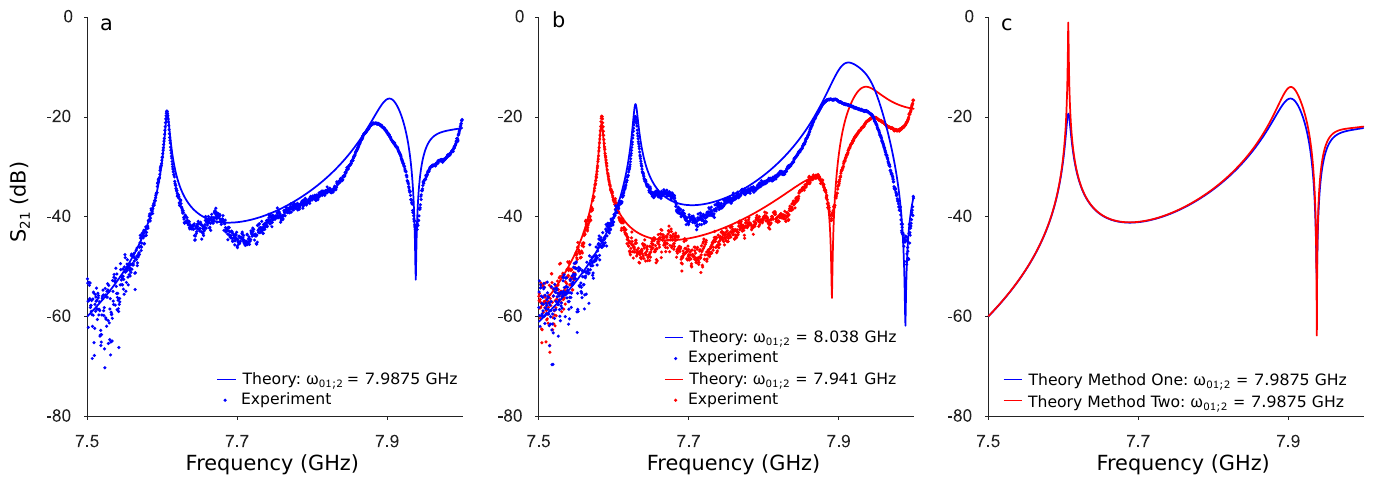}
\caption{Dependence of transmission $S_{21}$ on drive frequency, dependence used for determining the coupling strength of the second qubit, i.e.\ the one at site $N/2 + 1 = 9$. \textbf{a} Solid blue line is theoretical data for  $\omega_{01;2}=7.9875$ GHz while blue dots are experimental data. We choose parameters of the Hamiltonian such that the bound state frequency, the transmission dip frequency, and the linewidth of the bound state match the experimental data. \textbf{b} Solid blue line is theoretical data for $\omega_{01;2}=7.941$ GHz and blue dots are experimental data. Solid red line is theoretical data for $\omega_{01;2}=8.038$ GHz and red dots are experimental data. \textbf{c} Comparison of transmission methods for $\omega_{01;2}=7.9875$ GHz. The solid blue line is from method one, while the solid red line is from method two. The bound state and transmission dip occur at the same frequencies for both methods.}
\label{Transport}
\end{figure}

The first parameters we determine are  $\kappa$ and $\kappa_0$ (using the first transmission method). We turn off the coupling of the qubits to the photonic crystal (in the experiment, this is accomplished by saturating the qubits). We set $\kappa_0$ and $\kappa_q$ to zero and fit $\kappa$. Given that the largest losses occur at the ends of the photonic crystal, fitting $\kappa$ first is reasonable. $\kappa$ controls the linewidth of the photonic modes and the transmission amplitude difference between the transmission dips and peaks in the passband. We find that a reasonable estimate for $\kappa$ (for the experimental data in Fig.~\ref{crystal}d) is $1$ GHz, although any $\kappa$ in the range of $.5$ to $1.5$ GHz also gives a reasonable fit. We then turn on $\kappa_0$, which further reduces the transmission amplitude difference between the transmission dips and peaks in the passband and lowers the transmission peak of the lowest photonic mode. We estimate that $\kappa_0=4$ MHz (although any $\kappa_0$ in the range of $3$ MHz to $5$ MHz also fits the data well). Fig.~\ref{crystal}d shows that simulated transmission is in good agreement with the experimental data. Given that there are other losses in the system that we have not included, these numbers should be understood as estimates. 

We now turn to determining $g_i$. While $\kappa_q$ is set to zero for now, we find that varying it or the other loss parameters does not noticeably change the frequency of the bound state or transmission dip, thus making our estimate of the coupling strength independent of the decay parameters. To begin, we detune the qubit at site $N/2$ far away from the passband (in the experiment, the detuned qubit is at $4.5$ GHz) and then fix the other (i.e.\ second) bare qubit frequency \cite{Note1}. 
Transmission is then calculated as a function of driving frequency. We find that $g_2=.55$ GHz and $\omega_{01;2} =7.9875$ GHz match the experimental data well when the bound state is at $7.605$ GHz, as seen in Fig.~\ref{Transport}a. Calculating transmission when the first qubit frequency is near the pass band and the second qubit frequency is detuned and comparing it to experimental data, we find $g_1=.505$ \cite{Note1}. To make sure we have chosen suitable coupling strengths, we tune the bare qubit frequency (the detuned bare qubit frequency is kept fixed). If we have chosen the correct parameters, we should accurately capture the locations of the bound state and the transmission dip for different bare qubit frequencies, while keeping the other parameters fixed. Indeed, as seen in Fig.~\ref{Transport}b, we find this is the case for the chosen coupling strengths.

Our next goal is to obtain an estimate for $\kappa_q$. To do so, we increase $\kappa_q$, which increases the linewidth of both the transmission dip and the bound state, until the linewidth of the bound state matches the experimental value well for a fixed bare qubit frequency (we note increasing $\kappa_0$ also increases the linewidth of the bound state, however $\kappa_0$ is already fixed). We found that $\kappa_q=.5$ MHz accomplishes this task for $\omega_{01;2}=7.9875$. To make sure we have chosen a suitable qubit halfwidth, we again tune the second bare qubit frequency (while keeping the detuned bare qubit frequency fixed). If we have chosen a reasonable qubit halfwidth, we should be able to accurately estimate the linewidth of the bound state for different bare qubit frequencies (while keeping all other parameters fixed). Fig.~1b of the main text shows that our estimate of $\kappa_q$ is reasonable. 

Before moving on, we briefly comment on the second transmission method. Fig.~\ref{Transport}c shows theoretical data from both simulations for $g_2=.55$ GHz and $\omega_{01;2} = 7.9875$ GHz. The locations of the bound state and transmission dip occur at the same spot for both methods. The key difference is that the second method does not accurately predict the magnitude of the linewidth of the bound state as it assumes $\kappa_q = \kappa_0 = \kappa= 0$ (but instead has coherent coupling of the crystal to the waveguide). In these simulations, we have taken $g_w=2$ GHz as that fits the data when the qubits are saturated well (not shown). We note that any value of $g_w$ in the range of $1.5$ to $2.5$ GHz also gives a reasonable fit to the saturated qubit data and that the frequencies of the bound state and the transmission dip are not sensitive to $g_w$. Simulated transmission data presented in the main text is from method one.

We now fit the last remaining parameter, $\Delta$. We first diagonalize $H_{tot}$ in the two-excitation sector for fixed $\omega_{01;i}$ (with the other qubit detuned far away). The theoretical prediction for the dressed anharmonicity is found by taking the lowest eigenvalue of $H_{tot}$ in the two-excitation sector and subtracting two times the lowest single-excitation eigenvalue. We vary $\Delta$ until the theoretically predicted dressed anharmonicity matches the experimentally measured dressed anharmonicity for a given bare qubit frequency (we choose our bare qubit frequency such that the single-particle bound state is at $7$ GHz). In doing so, we find, to a good approximation, that $\Delta=.365$ GHz for both qubits. We then vary the bare qubit frequency and make sure the theoretically predicted dressed anharmonicity is still consistent with the experimentally measured value for different bare qubit frequencies. We find excellent agreement for a wide range of bare qubit values as shown in Figs.~2a and 2b in the main text.

Before we close this section, we estimate errors in our parameters. To begin, we estimate what change in hopping parameters (we call these different hopping parameters $J'$) we would get if we choose $Z_{high}=123.5 \Omega$ instead of $124\Omega$. Both of these choices fit the experimental data well in transmission matrix simulations. This choice of $Z_{high}$ gives $J'_0=9.331$ GHz, $J'_1=0.7308$ GHz, $J'_2=-0.0345$ GHz, $J'_3=0.0179$ GHz, $J'_4=-0.0035$ GHz, and $J'_5=0.0014$ GHz. We see that the ratio of these hopping parameters to the previous set is not less than .97 for any term, so we expect the other parameters in our model will not differ by more than 5 percent from their given predictions. We also expect this to hold true if one uses any reasonable set of photonic crystal parameters. To test this, we estimate $g_2$ and the range of decay parameters for the second set of hopping parameters. We find $g_2=.555$ GHz and that the same range of decay parameters fit the data well, consistent with our expectation.

\section{Bound State Fundamentals}

\begin{figure}
\centering
\includegraphics[scale=2]{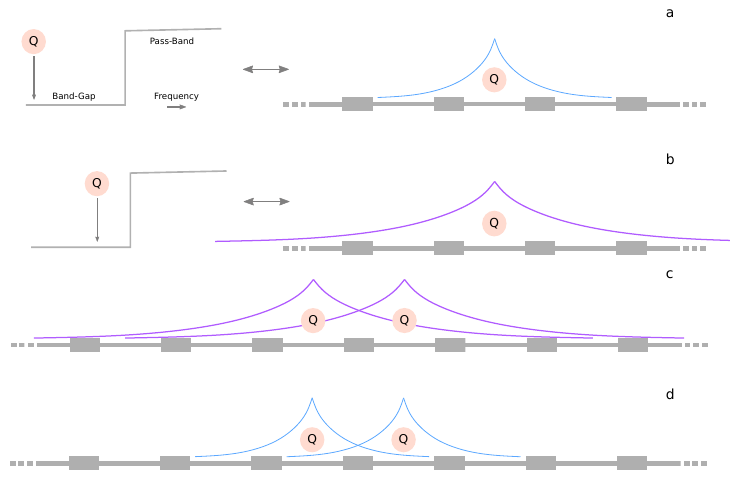}
\caption{Visualizing bound states - A qubit (pink circle) is coupled to one site of a 1D photonic crystal (gray line of alternating width). Coupling a qubit and band edge produces a photonic envelope (blue/purple) that is spatially centered at the qubit location. \textbf{a} and \textbf{b} The localization of the photonic component is determined by the detuning of the qubit from the band edge. \textbf{a} is more localized than \textbf{b} as the detuning is larger in the former. The overlap of the photonic wavefunction with the ends of the crystal (not shown) determines the linewidth of the bound state measured in transmission.  The strength of the interaction between the bound states, when qubits are resonant with one another, can be understood in \textbf{c} and \textbf{d} as depending on the localization of the bound states. }
\label{diagram}
\end{figure}
\begin{figure}
\centering
\includegraphics[scale=1]{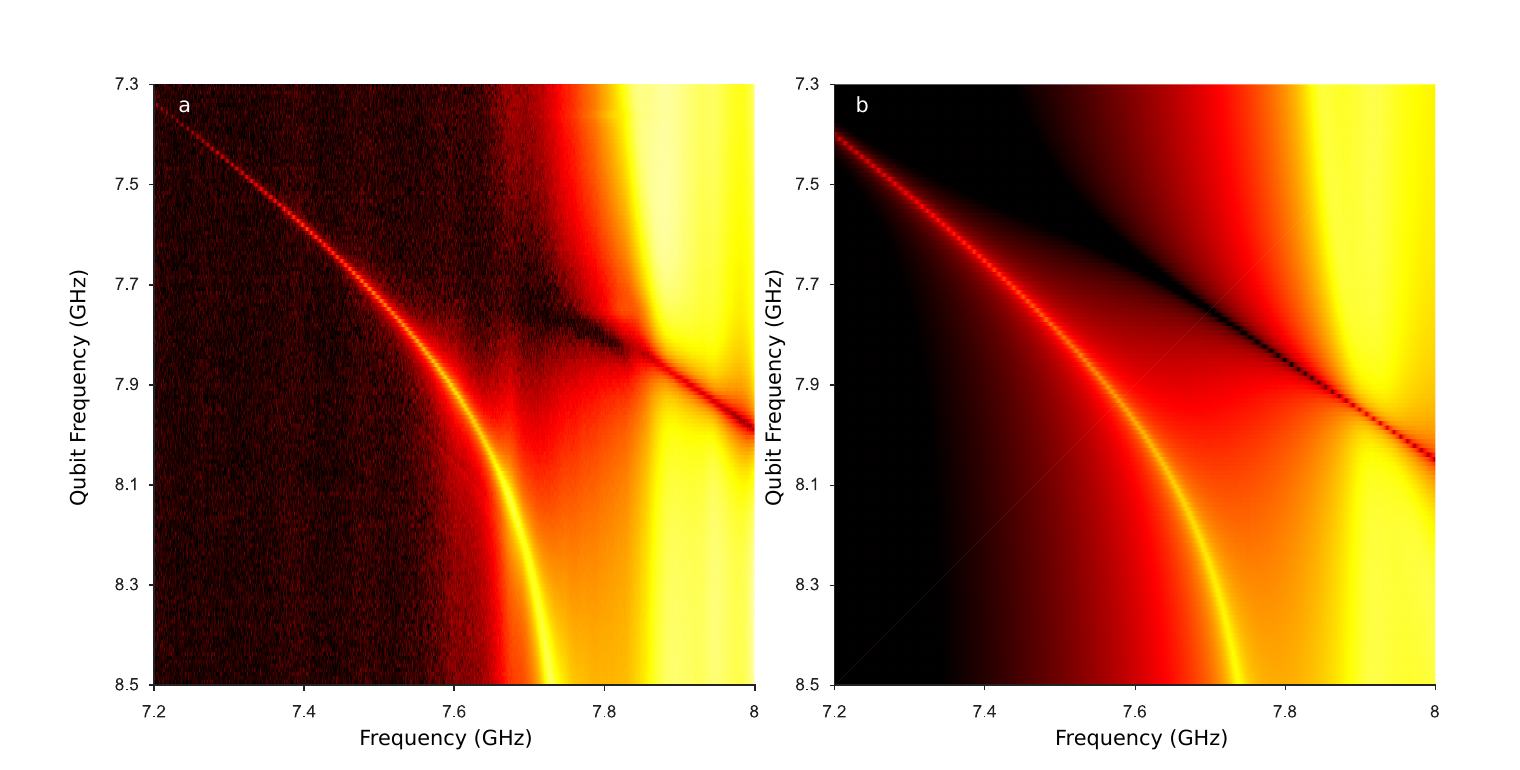}
\caption{\textbf{a} Experimental data and \textbf{b} hopping model simulation for $S_{21}$ vs.\ single-qubit frequency and probe frequency. The bare band edge is at 7.797GHz. The bright peak in the bandgap is the dressed qubit-photon bound state. The bound state always exists within the bandgap for qubit frequencies both above and below the band edge - a clear signature of non-Markovianity. In this figure, the other qubit is far detuned and has negligible effect.
}
\label{SingleQTune}
\end{figure}

	Strong light-matter interaction between atoms and slow-light structures, ones with vanishing or significantly reduced group velocity such as photonic band edges in photonic crystals, has been an area of ongoing interest both in theory and recent experiment. The principal interest behind this study is the localized \textit{bound} photonic state that forms around the atom. This bound state has an exponentially decaying photonic envelope that tunes with detuning of the qubit transition from the band edge (see Fig.~\ref{diagram}). While the bound state is always within the gap, the frequency of the bound state changes with qubit frequency (see Fig.~\ref{SingleQTune}). Additionally, the state becomes less localized as the bare qubit is tuned closer to the band edge (see Fig.~\ref{diagram}a,b). 

In a finite-size system, these localized states overlap with the ends of the crystal, thus facilitating single-photon transport across the crystal at the bound-state frequency and providing an avenue to probe these states. This tunable photonic interaction mechanism provides a platform for simulation of many-body quantum optics in one-dimensional systems, distinct from cavity or waveguide QED (see Fig.~\ref{diagram}c,d).

	In Fig.~\ref{SingleQTune} we measure $S_{21}$ at low power to track the bound state as a function of qubit frequency. The bound state shows up as a lorentzian peak in transmission. We detect the change in wavefunction overlap as a change in bound state linewidth - where linewidth increases with localization length (see Fig.~1c in Ref.\ \cite{Liu2016}).  As discussed in Sec.~\ref{expDetails}, the localization of the photonic wavefunction is determined predominantly by the strength of the coupling, properties of the band edge, and the frequency detuning between the atom and the band edge. The ability to tune the localization with  small sized crystals shows the versatility of this platform. 

 \subsection{Bound State Linewidth Dependence on Detuning }
	
	The linewidth of the bound state is set by the amplitude of the exponentially decaying photonic wavefunction at the ends of the crystal (ignoring other forms of loss). The envelope amplitude decays as $\sim e^{-x/L}, $ where x is the distance from the qubit location and L is the localization length. For simplicity, we consider the case of a single qubit at the center of the crystal (total length $d$) such that the envelope is symmetric. This results in a linewidth proportional to $e^{-d/2L}$. Of course, this approximation is only valid in the limit where $e^{-d/2L} <<  1$, meaning the bound state is sufficiently localized compared to the finite-length of the crystal. 
	
	
		 
\subsection{Single-Photon Bound States: Exact Solution}\label{sec:exacsol}
In this section, we discuss the theory of atom-photon bound states in the single-excitation sector for an infinite photonic crystal coupled to two qubits. While our system is of course finite, these results provide an intuitive understanding of our system. We note that a similar calculation for two qubits was done in Ref.~\cite{Calajo2016} for the case when the two qubits have equal coupling strengths and equal qubit frequencies. The results below generalize that work to unequal coupling strengths and unequal qubit frequencies. To begin, we first write the Fourier transformed Hamiltonian (ignoring decay and ignoring terms that do not affect the single-excitation bound states),
\begin{align}
H=\sum_{k}\omega_k a^\dagger_ka^{\phantom{\dagger}}_k+ \sum_{i=1,2}\omega_{01;i}|1\rangle_i\langle 1|_i+\frac{1}{\sqrt{N}}\sum_{i=1,2}\bigg(\sum_{k} g_i(a^\dagger_{k}e^{ikaz_i}|0\rangle_i\langle 1|_i+a^{\phantom{\dagger}}_{k}e^{-ikaz_i}|1\rangle_i\langle 0|_i)\bigg).
\end{align}
To make analytical progress, we assume that the dispersion relation is $\omega_k =\omega_0+\alpha a^2(k\mp \frac{\pi}{a})^2$, which is valid around $k=\pm \pi/a$.
While we have chosen a quadratic dispersion, these results are qualitatively similar for a cosine dispersion. The most general single-excitation wavefunction is
\begin{equation}
|\psi_{1B}\rangle=a_1|1\rangle_{1}|0\rangle_{2}|0\rangle+a_2|0\rangle_{1}|1\rangle_{2}|0\rangle+\sum_{k}c_{k}a^{\dagger}_k|0\rangle_{1}|0\rangle_{2}|0\rangle.
\end{equation}
Here, the basis states are labeled as $|qubit~one\rangle_1| qubit~two\rangle_2 |photon \rangle$. Solving the eigenvalue equation, $H|\psi_{1B}\rangle=E_{1B}|\psi_{1B}\rangle$, yields the following coupled equations,
\begin{equation}
E_{1B}a_1=\omega_{01;1}a_1+\frac{g_1}{\sqrt{N}}\sum_kc_ke^{-ikaz_1},
\label{aoneeq}
\end{equation}
\begin{equation}
E_{1B}a_2=\omega_{01;2}a_2+\frac{g_2}{\sqrt{N}}\sum_kc_ke^{-ikaz_2},
\end{equation}
\begin{equation} \label{eq:ck}
E_{1B}c_k=\omega_kc_k+a_1\frac{g_1}{\sqrt{N}}e^{ikaz_1}+a_2\frac{g_2}{\sqrt{N}}e^{ikaz_2}.
\end{equation}
Solving for $c_k$ from Eq.\ (\ref{eq:ck}) and inserting the result into Eq.~(\ref{aoneeq}) yields
\begin{equation}
E_{1B}a_1=\omega_{01;1}a_1+a_1\frac{g_1^2}{N}\sum_k\frac{1}{E_{1B}-\omega_k}+a_2\frac{g_1g_2}{N}\sum_k\frac{e^{-ika(z_1-z_2)}}{E_{1B}-\omega_k}.
\end{equation}
The sums are evaluated as follows,
\begin{align}
\frac{1}{N}\sum_k\frac{1}{E_{1B}-\omega_k}=a\int_0^{\pi/a}\frac{dk}{2\pi}\frac{1}{E_{1B}-\omega_0-\alpha a^2(k-\frac{\pi}{a})^2}+a\int_{-\pi/a}^{0}\frac{dk}{2\pi}\frac{1}{E_{1B}-\omega_0-\alpha a^2(k+\frac{\pi}{a})^2}.
\end{align}
Shifting the integrals by $\pi/a$, making the integrals dimensionless, and extending the limits to infinity gives,
\begin{align}
\frac{1}{N}\sum_k\frac{1}{E_{1B}-\omega_k}=\int_{-\infty}^{\infty}\frac{d\tilde{k}}{2\pi}\frac{1}{E_{1B}-\omega_0-\alpha \tilde{k}^2}=-\frac{1}{2\sqrt{\alpha(\omega_0-E_{1B})}}.
\end{align}
Here, we have assumed that $E_{1B}<\omega_0$. Following the same steps for the remaining integral gives
\begin{align}
\frac{1}{N}\sum_k\frac{e^{-ika(z_1-z_2)}}{E_{1B}-\omega_k}=
-\frac{\cos(\pi(z_1-z_2))}{2\sqrt{\alpha(\omega_0-E_{1B})}}e^{-\sqrt{\frac{\omega_0-E_{1B}}{\alpha}}|z_1-z_2|}.
\end{align}
Here, we used the fact that $(z_1 - z_2)$ is an integer. We then have
\begin{equation}
E_{1B}a_1=\omega_{01;1}a_1-\frac{1}{2\sqrt{\alpha(\omega_0-E_{1B})}}\bigg(a_1g_1^2+a_2g_1g_2\cos(\pi(z_1-z_2))e^{-\sqrt{\frac{\omega_0-E_{1B}}{\alpha}}|z_1-z_2|}\bigg).
\end{equation}
Repeating these steps for $a_2$, gives the following equation for the bound state energy,
\begin{align}
\bigg(E_{1B}-\omega_{01;1}+\frac{g_1^2}{2\sqrt{\alpha(\omega_0-E_{1B})}}\bigg)\bigg(E_{1B}-\omega_{01;2}+\frac{g_2^2}{2\sqrt{\alpha(\omega_0-E_{1B})}}\bigg)=\frac{g_1^2g_2^2e^{-2\sqrt{\frac{\omega_0-E_{1B}}{\alpha}}|z_1-z_2|}}{4\alpha(\omega_0-E_{1B})},
\label{TWOBSGEN}
\end{align}
where we used the fact that $(z_1 - z_2)/a$ is an integer again.
We note that when the qubits are infinitely far away from each other, we recover the well-known bound state energy for a single qubit (see, for example, Ref.~\cite{Calajo2016} or Ref.~\cite{Liu2016}),
\begin{align}
E_{1B}-\omega_{01}=-\frac{g^2}{2\sqrt{\alpha(\omega_0-E_{1B})}}.
\label{BSENERGYKNOWN}
\end{align}
Generically, Eq.~(\ref{TWOBSGEN}) yields one or two bound states depending on the coupling strength, qubit frequencies, distance between qubits and $\alpha$ \cite{Calajo2016}.

To illustrate this point, we consider the case when $g_1=g_2=g$ and $\omega_{01;1}=\omega_{01;2}=\omega_{01}$ (which is relevant to the case in Fig.~3e of the main text). While the coupling strengths are not exactly equal in our experimental system, it is a decent approximation to consider them equal. In this case, we expect a symmetric and an antisymmetric solution, i.e. $a^e_{1}=a^e_{2}$ or $a^{o}_{1}=-a^{o}_{2}$. The difference of the bound-state energy and the energy of the band edge, $E_{1B}-\omega_0=\delta E_{1B}<0$, is then given by
\begin{equation}
\bigg(\delta E_{1B}-(\omega_{01}-\omega_0)\bigg)=\Sigma_{\pm}(\delta E_{1B})=-\frac{g^2}{2\sqrt{-\alpha\delta E_{1B}}}\bigg(1\pm (-1)^{|z_1-z_2|}e^{-\sqrt{\frac{-\delta E_{1B}}{\alpha}}|z_1-z_2|}\bigg),
\label{sym_exact}
\end{equation}
 where $+$ is for the symmetric state and $-$ is for the antisymmetric state and $\Sigma_{\pm}(\delta E_{1B})$ is the self-energy. The condition for the presence of a bound state, as derived in Ref. ~\cite{Calajo2016} is $-(\omega_{01}-\omega_0)>\Sigma_{\pm}(0)$.

We explicitly consider the experimentally relevant case when $|z_1-z_2|$ is odd. In this case, we have $\Sigma_{+}(0)=-\frac{g^2}{2\alpha}|z_1-z_2|$ and $\Sigma_{-}(0)=-\infty$. For the antisymmetric state, the condition is always satisfied, while for the symmetric state, we only have a bound state if $g>\sqrt{\frac{2\alpha(\omega_{01}-\omega_0)}{|z_1-z_2|}}$. We now apply this formalism to our experimental system. For our experimental system, $\alpha=1.155$ GHz and $|z_1-z_2|=1$. We also take $g$ to be the average of the two coupling strengths determined in the previous section, i.e. $g=(g_1+g_2)/2 =.5275$ GHz. Using these numbers, we estimate that we have two bound states for $\omega_{01}-\omega_0<120$ MHz. We remind the reader that this result is only an estimate, as our experimental system is finite and has unequal coupling strengths.

Finally, one can investigate the real-space wave function. Fourier transforming $c_k$ gives
 \begin{align}
 c_j\propto \frac{1}{N}\sum_k c_ke^{ikax}=a_1g_1\int_{-\infty}^\infty  \frac{dk}{2\pi}\frac{e^{ika(j-z_1)}}{E_{1B}-\omega_k}+a_2g_2\int_{-\infty}^\infty \frac{dk}{2\pi} \frac{e^{iak(j-z_2)}}{E_{1B}-\omega_k}=\nonumber \\
 -a_1g_1\frac{\cos(\pi(j-z_1))}{2\sqrt{\alpha(\omega_0-E_{1B})}}e^{-\sqrt{\frac{\omega_0-E_{1B}}{\alpha}}|j-z_1|}-a_2g_2\frac{\cos(\pi(j-z_2))}{2\sqrt{\alpha(\omega_0-E_{1B})}}e^{-\sqrt{\frac{\omega_0-E_{1B}}{\alpha}}|j-z_2|}.
 \end{align}
We see that the photon is exponentially localized around the qubits with localization length $a \sqrt{\frac{\alpha}{\omega_0-E_{1B}}}$, as in the case of a single qubit (see Fig.~\ref{diagram}). We remind the reader that we can have two different bound state energies (one for the symmetric bound state and one for the antisymmetric one), thus two different localization lengths. In other words, the linewidths of the bound states will, in general, be different, with the bound state closer to the band edge having a larger linewidth. In our system, the symmetric bound state, whenever it exists, is closer to the band edge and thus has a larger linewidth. 

\subsection{Single-Photon Bound States: Born-Markov Solution}
\begin{figure}
\centering
\includegraphics[scale=1.1]{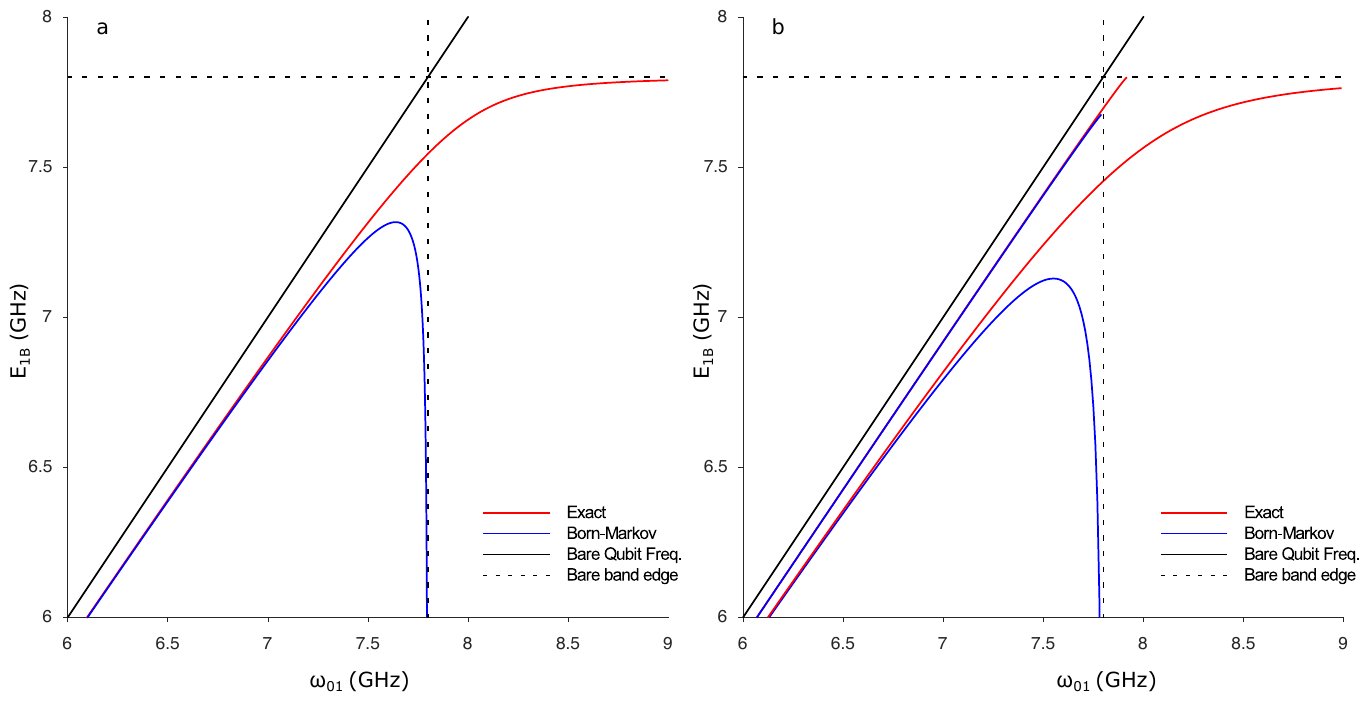}
\caption{Exact solution versus Born-Markov Approximation. Here, the red lines are the exact solution while the blue lines are the Born-Markov approximation. The dashed black lines mark the band edge and the thick black line is the bare qubit frequency. Here we take, $\omega_0=7.8$ GHz, $\alpha=1.155$ GHz and $g=.5275$ GHz. \textbf{a} Single qubit. \textbf{b} Two qubits for $|z_1-z_2|=1$.}
\label{Markov}
\end{figure}
We now briefly compare these exact results to the Born-Markov approximation. For simplicity, we restrict ourselves to the case when the qubit frequencies and coupling strengths are the same. Using a second-order Schrieffer-Wolff transformation to eliminate the high-energy subspace gives the following effective Hamiltonian for the two qubits,
\begin{align}
H_{BM}=\bigg(\omega_{01}+
\frac{g^2}{N}\sum_k\frac{1}{\omega_{01}-\omega_k}\bigg)\bigg(|1\rangle_1\langle 1|_1+|1\rangle_2\langle 1|_2\bigg)+\nonumber \\
\frac{g^2}{N}\sum_k\frac{e^{ika(z_1-z_2)}}{\omega_{01}-\omega_k}\bigg(|0\rangle_1\langle 1|_1|1\rangle_2\langle 0|_2+|1\rangle_1\langle 0|_1|0\rangle_2\langle 1|_2\bigg)=\nonumber \\
\bigg(\omega_{01}-\frac{g^2}{2\sqrt{\alpha(\omega_0-\omega_{01})}}\bigg)\bigg(|1\rangle_1\langle 1|_1+|1\rangle_2\langle 1|_2\bigg)-\nonumber \\
\frac{g^2}{2\sqrt{\alpha(\omega_0-\omega_{01})}}\cos(\pi(z_1-z_2))e^{-\sqrt{\frac{\omega_0-\omega_{01}}{\alpha}}|z_1-z_2|}\bigg(|0\rangle_1\langle 1|_1|1\rangle_2\langle 0|_2+|1\rangle_1\langle 0|_1|0\rangle_2\langle 1|_2\bigg).
\label{BornMARHAM}
\end{align}
We stress that this formula is only valid for $\omega_{01}<\omega_{0}$.  
Diagonalizing $H_{BM}$, we have the following eigenvalues,
\begin{equation} \label{eq:E1}
E_1=\omega_{01}-\frac{g^2e^{-\frac{1}{2}\sqrt{\frac{\omega_0-\omega_{01}}{\alpha}}|z_1-z_2|}}{\sqrt{\alpha(\omega_0-\omega_{01})}}\sinh\bigg(\frac{1}{2}\sqrt{\frac{\omega_0-\omega_{01}}{\alpha}}|z_1-z_2|\bigg),
\end{equation}
\begin{equation}
E_2=\omega_{01}-\frac{g^2e^{-\frac{1}{2}\sqrt{\frac{\omega_0-\omega_{01}}{\alpha}}|z_1-z_2|}}{\sqrt{\alpha(\omega_0-\omega_{01})}}\cosh\bigg(\frac{1}{2}\sqrt{\frac{\omega_0-\omega_{01}}{\alpha}}|z_1-z_2|\bigg).
\end{equation}
In the limit where the qubits are infinitely far apart, we recover the standard expression for the dressed qubit frequency,
\begin{align}
\omega_{01}'=\omega_{01}-\frac{g^2}{2\sqrt{\alpha(\omega_0-\omega_{01})}}.
\end{align}		 
In Fig.~\ref{Markov}, we compare the results obtained in the Born-Markov approximation to the exact analytical results. Fig.~\ref{Markov}a shows these results for a single qubit and Fig.~\ref{Markov}b shows these results for two qubits with $|z_1-z_2| = 1$. As expected, the Born-Markov approximation is a good approximation when the qubit frequency is away from the band edge. Closer to the band edge, the Born-Markov approximation fails, particularly for the lower-energy (i.e. antisymmetric) state. Furthermore, by comparing the blue curve in Fig.~\ref{Markov}a to the experimentally measured bound state in Fig.~\ref{SingleQTune}, we clearly see that  the experiment is not well-described by the Born-Markov approximation. We note that the higher energy red line in Fig.~\ref{Markov}b ends abruptly at $\omega_{01}\approx7.920$ GHz as there is only one bound state for $\omega_{01}>\omega_{0}+.120$ GHz. 

We now analytically show that the Born-Markov approximation is excellent for one of the dressed states close to the band edge. We begin by expanding Eq.~(\ref{sym_exact}) (for the symmetric case, when $|z_1-z_2|=1$) in the limit that $\omega_0-E_{1B}\ll\alpha$, i.e. when the bound-state energy is close the band edge. This gives
\begin{equation}
E_{1B}=\omega_{01}-\frac{g^2}{2\alpha}(1+O\bigg(\frac{E_{1B}-\omega_0}{\alpha}\bigg)).
\end{equation}
Now comparing to the Born-Markov solution for $E_1$ in Eq.\ (\ref{eq:E1}) around $\omega_0\approx\omega_{01}$, we have $ E_1\approx\omega_{01}-\frac{g^2}{2\alpha}$. We thus see that one of the dressed states (the symmetric one) is well captured by the Born-Markov approximation, while the other is not as seen in Fig.~\ref{Markov}b.

\subsection{Single-Photon Transport via the Bound State}
    As discussed, a bound state mediates transport across the crystal, at the otherwise forbidden frequencies inside the bandgap, via the overlap of the photonic mode with the ends of the crystal. However, unlike a cavity mode which accommodates many photons (of the same frequency) due to its harmonic nature, the bound state inherits an anharmonic level structure from the qubit. This will result in single-photon, blockaded transport. For a definitive confirmation, we measure the second-order autocorrelation of the transmitted component of a weak, resonant, continuous drive.
    
\begin{figure}
\centering
\includegraphics[scale=1]{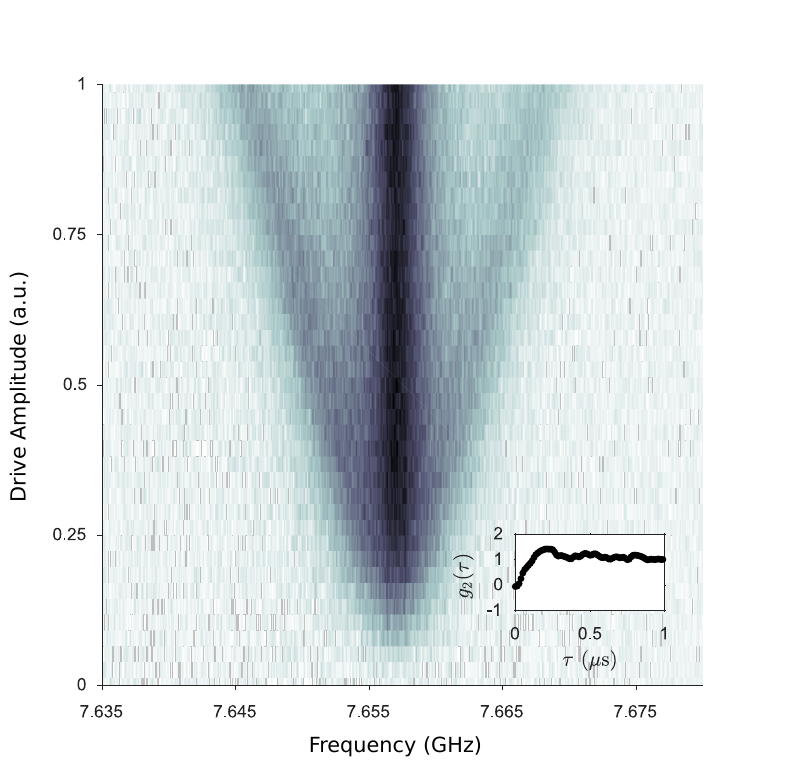}
\caption{Power spectrum of a resonantly driven bound state for increasing drive amplitude. Sidebands are linearly displaced from the central peak with increasing drive amplitude, characteristic of the Mollow triplet. Inset - second order auto-correlation measurement for drive amplitude = 0.2 is consistent with single photon, anti-bunched transport.}
\label{g2}
\end{figure}

    In Fig.~\ref{g2}, we plot the emission spectrum of the resonantly driven bound state for low drive amplitudes. Here we see the familiar Mollow triplet structure featuring sidebands that are linearly displaced from the center peak with increasing drive amplitude. We measure second-order autocorrelation (Fig.~\ref{g2} inset) for a drive amplitude below the threshold for incoherent triplet emission such that the qubit is not saturated by the drive. This measurement (see \cite{DaSilva2010,Eichler2011,Lang2011,Bozyigit2010,Hoi2012} for concept) was made possible by a TWPA (MIT Lincoln Labs) to improve SNR and a GPU (CUDA-Matlab) for significant computational speed-up. 
        	  
\section{Emission Theory}    
\begin{figure}
 \includegraphics[scale = 1.2]{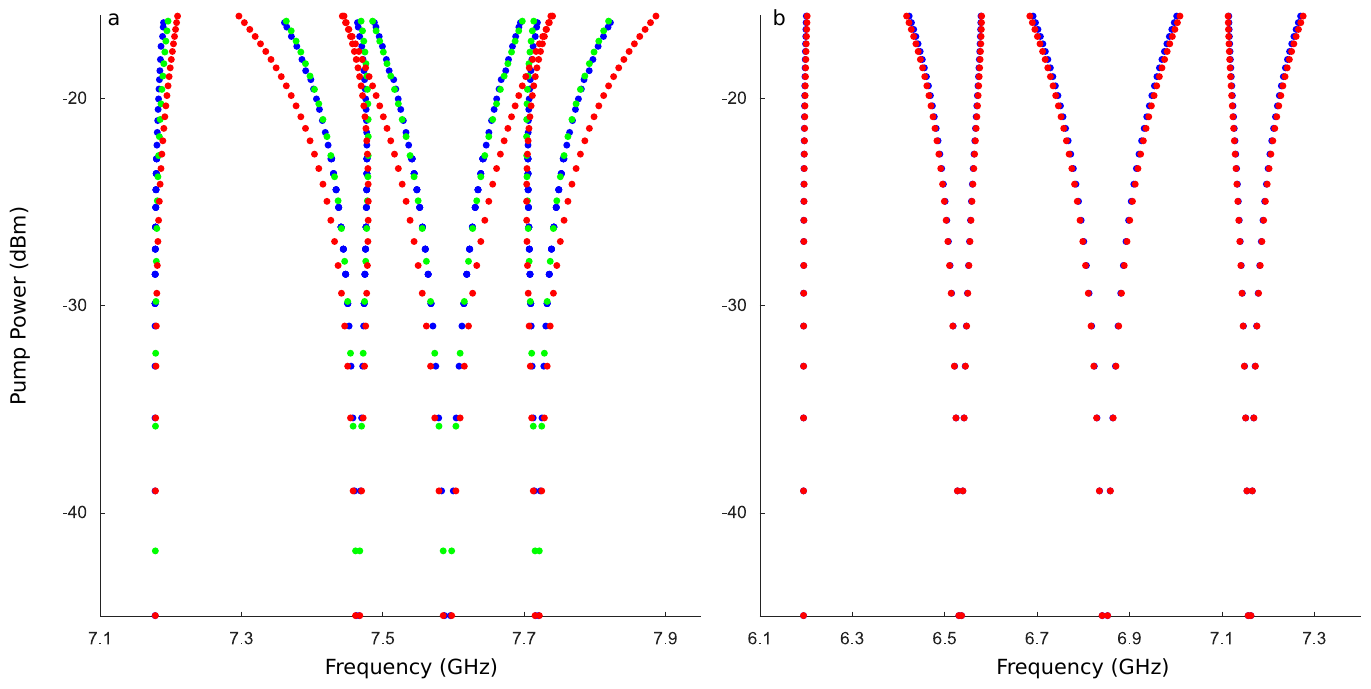}
\caption{Theoretical simulations of the emission spectrum for different pump powers. Here, the blue squares are numerical results obtained from the total Hamiltonian Eq.\ \eqref{Hamwdrive}, the green squares are from the dressed-qubit Hamiltonian Eq.\ \eqref{spinpump}, and the red squares are from Eq.\ \eqref{spinpump} but with $\cos \theta = 1$. \textbf{a} Here, $\omega_{01}=7.97$ GHz, and the bound-state frequency is $7.591$ GHz. Using the appropriately reduced matrix element ($\cos \theta < 1$, blue squares) is crucial in obtaining accurate results using Eq.~\eqref{spinpump}. \textbf{b} Here, the bare qubit frequency of $7$ GHz and the bound-state frequency of $6.847$ GHz are both far from the band edge, so that $\cos \theta \approx 1$ and the reduction in the matrix element can be neglected, so all three data sets lie on top of each other.}
\label{pump}
\end{figure}
In this section, we discuss the theoretical modeling of the emission spectrum of a resonantly driven bound state. Unfortunately, due to the large dimension of the Hilbert space, we found that a direct approach of calculating the emission spectrum using the master equation for our sixteen-unit-cell system is not numerically feasible. Instead, we diagonalize the Hamiltonian and investigate energy differences. While this approach does not predict the widths and the driving-strength-dependent intensities of the sidebands, it does predict the frequencies of the sidebands. The Hamiltonian of a single qubit with a driving frequency $\omega_d$ equal to the bound-state frequency is given by (in the rotating frame) \cite{Note2} 
\begin{align}
H=\sum\limits_{i,j}(J_{i,j}-\omega_d\delta_{i,j})a^\dagger_i a^{\phantom{\dagger}}_{j}+g_2\big(a^\dagger_{z_2}(|0\rangle\langle 1|+\sqrt{2}|1\rangle\langle 2|+\sqrt{3}|2\rangle\langle 3|+\sqrt{4}|3\rangle\langle 4|)+h.c.\big)+\nonumber \\
\sum_{n=0}^4(\omega_{0n}-n\omega_d)|n\rangle\langle n|+\Omega\big((|0\rangle\langle 1|+\sqrt{2}|1\rangle\langle 2|+\sqrt{3}|2\rangle\langle 3|+\sqrt{4}|3\rangle\langle 4|)+h.c.\big),
\label{Hamwdrive}
\end{align}
where $\Omega$ is the bare Rabi frequency of the drive and is the only unknown parameter. We stress that $\omega_d$ is not at the bare qubit frequency but at the frequency of the bound state. In the presence of a drive, excitation number is no longer conserved, and thus, to make progress, one must implement a cut-off. In our numerical simulations, we have implemented a cut-off of five qubit levels and 3 photons. Diagonalizing the system, we take the differences of the eigenvalues of states corresponding to large occupation of atomic states with no photons.

We now compare the results obtained by diagonalizing Eq.~\eqref{Hamwdrive} to the results obtained by driving a dressed qubit (without explicitly including the photonic crystal). The latter approach was used in Ref.~\cite{Liu2016}. The Hamiltonian for the dressed qubit is
\begin{align}
H=\sum_{n=0}^4(\tilde{\omega}_{0n}-n\omega_d)|n\rangle\langle n|+\tilde{\Omega}\bigg(|0\rangle\langle 1|+\sqrt{2}|1\rangle\langle 2|+\sqrt{3}|2\rangle\langle 3|+\sqrt{4}|3\rangle\langle 4|+h.c.\bigg).
\label{spinpump}
\end{align}
Here, $\tilde{\omega}_{0n}$ are the dressed qubit frequencies, $\omega_d=\tilde{\omega}_{01}$, and $\tilde{\Omega}$ is the Rabi frequency seen by the dressed qubit. If the exact wavefunction for the bound state is $|\psi\rangle=\cos\theta|1\rangle |0\rangle+\sin\theta\sum_{k}c_ka_k^\dagger|0\rangle|0\rangle$ with $\sum_k|c_k|^2=1$ (here, our basis states are labeled as $|qubit \rangle| photon\rangle $), we have $\tilde{\Omega}\approx \Omega \cos\theta$, where $\theta$ is given by
\begin{equation}\label{eq:theta}
\tan^2\theta=\frac{g^2}{N}\sum_{k}\frac{1}{(E_{1B}-\omega_k)^2},
\end{equation}
which, for an infinite system, simplifies to
\begin{equation}
\tan^2\theta=\frac{g^2}{4}\frac{1}{(E_{1B}-\omega_0)^{3/2}\sqrt{\alpha}}.
\end{equation}
When the bound state is approximately at $7.59$ GHz, we find $\cos\theta\approx.68$ for our finite system. Fig.~\ref{pump} compares the results obtained from Eqs.\ \eqref{Hamwdrive} and \eqref{spinpump}. We see that, upon taking into account the reduction of the matrix element due to the dressing, the two methods agree. We also see (Fig.\ \ref{pump}b) that, as expected, as the bare qubit frequency moves deeper into the gap, $\theta$ approaches zero and $\tilde \Omega$ approaches $\Omega$.  In Fig.~\ref{pump}, we have used the anharmonicity value predicted by theory. In Fig.~2c of the main text, we use the experimental values of anharmonicity, which is given by  $\tilde{\omega}_{02}-2\omega_d=-.11$ GHz. 

We assume that the feature around $7.22$ GHz in Fig. 2c of the main text is approximately $\omega_{23}$ (which gives $\tilde{\omega}_{03}-3\omega_{d}=-.48$ GHz). This assumption is in decent agreement (around $50$ MHz off) with results obtained by diagonalizing the full system when the bound state is at  $7.59$ GHz. This $50$ MHz disagreement can be traced back to the $20$ MHz disagreement in $\Delta$, the anharmonicity, between theory and experiment when the bound state is at  $7.59$ GHz (if $\Delta$ is off by $20$ MHz, level $|3\rangle$ is expected to be off from its value by around 3 times this amount as $\tilde{\omega}_{03}\approx 3\tilde{\omega}_{01}-3\Delta$). Finally, we find that $\tilde{\omega}_{04}-4\omega_d=-1.78$ GHz (this value is also consistent with results obtained by diagonalizing Eq.~\eqref{Hamwdrive}) matches the experimental data well as seen by overlaying the theoretical data from the dressed qubit with the experiential data (Fig. 2c of main text). In particular, we have captured the feature that appears around $7.22$ GHz and $-10$ dB. The unique bending of this feature can be traced back to the fact that the level structure of the dressed qubit [Eq.~\eqref{spinpump}] does not behave like a normal transmon due to the strong coupling to the photonic crystal.

\section{Multiphoton theory}
\subsection{Two-Photon Bound State}

\begin{figure}
\centering
\includegraphics[scale=1.1]{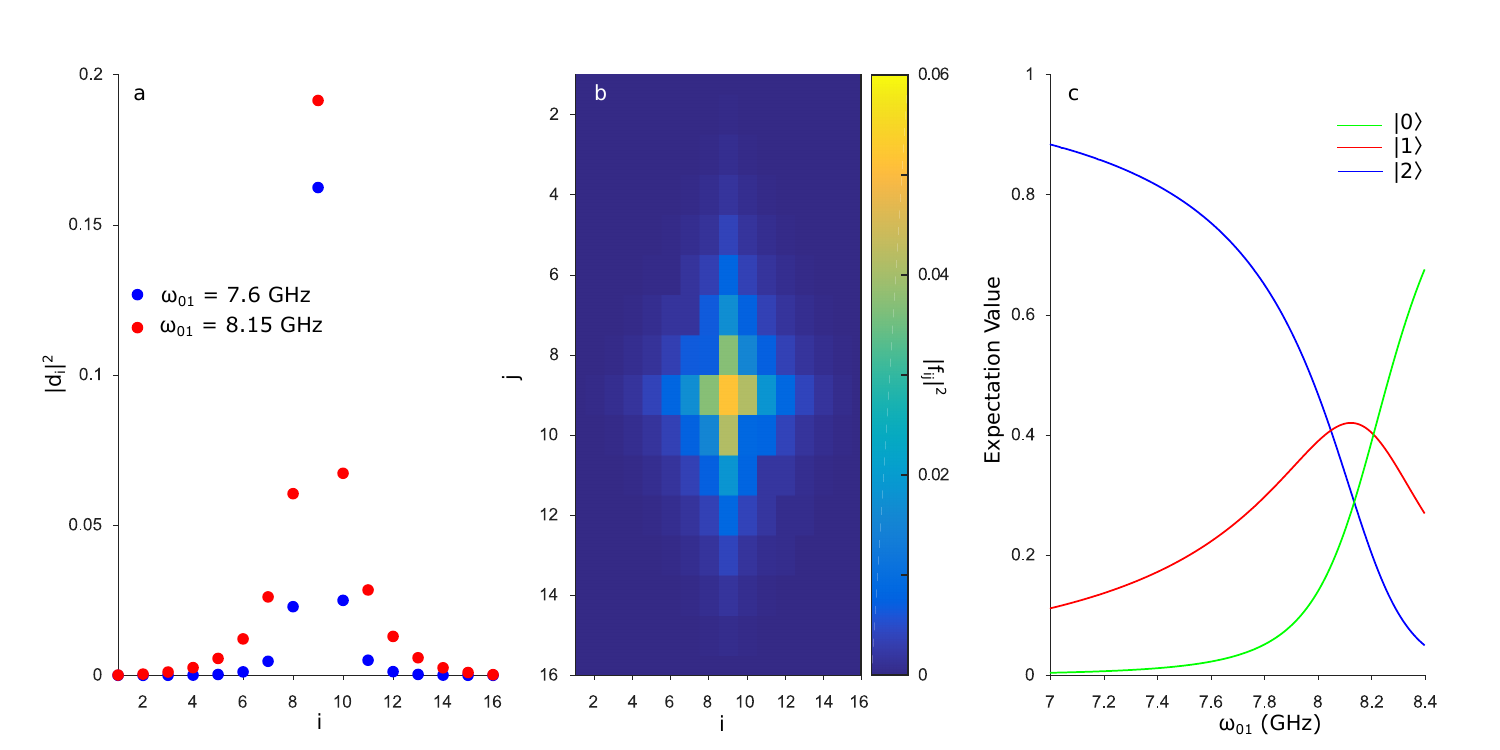}
\caption{Visualizing the two photon bound state. Here, the qubit is on site nine. \textbf{a} The blue dots are $|d_i|^2$ versus position for $\omega_{01}=7.6$ GHz. The red dots are $|d_i|^2$ for $\omega_{01}=8.15$ GHz as a function of position. \textbf{b} Plot of $|f_{i,j}|^2$ as a function of $i$ and $j$ for $\omega_{01}=8.15$ GHz. \textbf{c} The population of qubit levels in the two-excitation ground state of $H_{tot}$ as a function of bare qubit frequency.}
\label{Multiphoton}
\end{figure}

In this section, we discuss the two-photon bound state. For a single qubit, the most general two-excitation wave function is
\begin{equation}
|\psi_{2B}\rangle=b|2\rangle|0\rangle+\sum_{i}d_{i}a^\dagger_{i}|1\rangle|0\rangle+\sum_{i>j}f_{i,j}a^\dagger_{i}a^\dagger_{j}|0\rangle|0\rangle+\sum_{i}f_{i,i}\frac{(a^\dagger_{i})^2}{\sqrt{2}}|0\rangle|0\rangle.
\label{MPWF}
\end{equation}
Here the basis states are labeled as $|qubit\rangle|photon\rangle$. For $i < j$, it is convenient to define $f_{i,j}=f_{j,i}$. In Fig.~\ref{Multiphoton}, we plot $|d_{i}|^2$ and $|f_{i,j}|^2$, which are obtained via exact diagonalization of the two-excitation sector for 16 sites \cite{Note3}.
We observe that the photons are localized around the qubit. In Fig.~\ref{Multiphoton}c, we plot the populations of qubit levels in the two-excitation ground state of $H_{tot}$, which are given by
\begin{equation}
|\langle 0|\psi_{2B}\rangle|^2=|b|^2,~~~|\langle 1|\psi_{2B}\rangle|^2=\sum_{i}|d_i|^2,~~~~|\langle 2|\psi_{2B}\rangle|^2=\sum_{i\geq j}|f_{i,j}|^2,
\end{equation}
to illustrate which terms in Eq.~\eqref{MPWF} are important for a given bare qubit frequency. For example, when the bare qubit frequency is deep in the gap, the ground state of the two-excitation sector is mostly in the $|2\rangle|0\rangle$ state as seen in Fig.~\ref{Multiphoton}c. Upon increasing the bare qubit frequency (while still in the band gap), the population of $|1\rangle$ increases, while the population of $|0\rangle$ stays relatively small. This is because two photons must be exchanged to couple the dominate $|2\rangle|0\rangle$ state and any two-photon state and there are no terms in $H_{tot}$ that directly exchange two photons, thus making it a higher order process. On the other hand, coupling $|2\rangle|0\rangle$ and $a_{i}^\dagger|1\rangle|0\rangle$ only requires exchanging one photon. When the bare qubit frequency is at or near the band edge, each qubit level in Eq.~\eqref{MPWF} contributes significantly to the bound-state wavefunction.

Unfortunately, we are unaware of an exact solution similar to the one in Sec.~\ref{sec:exacsol}. To make analytical progress, we assume $f_{i,j}=0$ which is a valid approximation when the bare qubit frequency is deep in the band gap (as seen in Fig.~\ref{Multiphoton}c). Solving the eigenvalue equation, $H|\tilde{\psi}_{2B}\rangle=E_{2B}|\tilde{\psi}_{2B}\rangle$, for the following wave-function ansatz (in momentum-space),
\begin{equation}
|\tilde{\psi}_{2B}\rangle=\tilde{b}|2\rangle|0\rangle+\sum_{k}\tilde{d}_{k}a^\dagger_{k}|1\rangle|0\rangle,
\end{equation}
yields the following equations,
\begin{eqnarray}
\tilde{b}E_{2B}&=&\omega_{02}\tilde{b}+\frac{\sqrt{2}g}{\sqrt{n}}\sum_ke^{ikaz_2}\tilde{d}_k,\\
\tilde{d}_k E_{2B}&=&(\omega_{k}+\omega_{01})\tilde{d}_k+\frac{\sqrt{2}g}{\sqrt{N}}e^{-ikaz_2}\tilde{b}.
\end{eqnarray}
These are similar to equations for the single photon case. Thus, we have 
\begin{equation}
E_{2B}-\omega_{02}=-\frac{g^2}{\sqrt{\omega_0+\omega_{01}-E_{2B}}},
\end{equation}
and
\begin{equation}
d_j\propto \frac{\sqrt{2}\tilde{b}}{N}\sum_k\frac{e^{ika(j-z_2)}}{E_{2B}-\omega_{01}-\omega_k}=-\frac{\sqrt{2}\tilde{b}\cos(\pi(j-z_2))}{2\sqrt{\alpha(\omega_0+\omega_{01}-E_{2B})}}e^{-\sqrt{\frac{\omega_0+\omega_{01}-E_{2B}}{\alpha}}|j-z_2|}.
\end{equation}
We see that the photon is localized around the qubit, consistent with the exact finite-size numerical results seen in Fig.~\ref{Multiphoton}a. 
We note this ansatz breaks down when the bare qubit frequency is near the passband as the states become more photonic, in which case we can no longer neglect $f_{i,j}$.

\subsection{Two-Photon Avoided Crossing}
In this section, we discuss the two-photon avoided crossing seen in Fig.~4 of the main text. As in previous sections, we assume the two qubits have equal coupling strengths and equal bare frequencies. To begin, we first write the total Hamiltonian in the two-excitation sector (in the rotating frame), using a different notation,
\begin{equation}
H_2=H_{0x}+H_{0y}+H_{1x}+H_{1y},
\end{equation}
where
\begin{eqnarray}
H_{0x}&=&\sum_k(\omega_k+\omega_{12})(|0,1;k\rangle\langle 0,1;k|+|1,0;k\rangle\langle 1,0;k|),\\
H_{0y}&=&\sum_{k,p}(\omega_k+\omega_p-\omega_{02})|0,0;k,p\rangle\langle 0,0;k,p|,\\
H_{1x}&=&\frac{\sqrt{2} g}{\sqrt{N}}\sum_{k}(e^{ikaz_1}|1,0;k\rangle\langle 2,0;0|+e^{ikaz_2}|0,1;k\rangle\langle 0,2;0|+h.c.),\\
H_{1y}&=&\frac{g}{\sqrt{N}}\sum_{k,p;k\neq p}\bigg(e^{ipaz_1}|0,0;k,p\rangle\langle 1,0;k|+e^{ipaz_2}|0,0;k,p\rangle\langle 0,1;k|+h.c.)+\nonumber \\
&&
\frac{\sqrt{2} g}{\sqrt{N}}\sum_{k}\bigg(e^{ikaz_1}|0,0;k,k\rangle\langle 1,0;k|+e^{ikaz_2}|0,0;k,k\rangle\langle 0,1;k|+h.c.\bigg).
\end{eqnarray}
Here the basis states are labeled as $|qubit_1,~qubit_2;~photonic~field\rangle$. To proceed, we neglect the second line in $H_{1y}$ as there are many photonic modes, thus the probability of both photons going into the same mode is negligible. Using a unitary Schrieffer-Wolff transformation, we find the fourth order term in coupling strength, $g$, to be
\begin{equation}
H_{4}=\frac{1}{2}H_{1x}\tilde{H}_0(\tilde{H}_0H_{1x}^2+H_{1x}^2\tilde{H}_0)\tilde{H}_0H_{1x}-H_{1x}\tilde{H}_0H_{1y}\tilde{H}_0H_{1y}\tilde{H}_0H_{1x},
\label{fourthorder}
 \end{equation} 
 where $H_{0}=H_{0x}+H_{0y}$ and $\tilde{H}_0=H_0^{-1}$ (here, $H_0^{-1}$ is taken to be zero outside the support of $H_0$). We are only interested in terms that involve interactions between the $|0,2;0\rangle $ and $|2,0;0\rangle$ states, i.e. terms like $|2,0;0\rangle\langle 0,2;0|$. Only the last term in Eq.~(\ref{fourthorder}) contributes such a term. Ignoring contributions of the last term that are diagonal in the $\{|0,2;0\rangle,|2,0;0\rangle\}$ basis, the effective interaction between the $|2,0\rangle$ and the $|0,2\rangle$ states is (after projecting out the photonic degrees of freedom),
 \begin{align}
 H_{|2,0\rangle\leftrightarrow|0,2\rangle}=-\frac{4g^4}{N^2}|2,0\rangle\langle 0,2|\sum_{k,p}\frac{e^{i(k+p)a(z_2-z_1)}}{(\omega_k+\omega_p-\omega_{02})(\omega_k-\omega_{12})}\bigg(\frac{1}{\omega_{k}-\omega_{12}}+\frac{1}{\omega_{p}-\omega_{12}}\bigg)+h.c.=\nonumber \\
 -4g^4|2,0\rangle\langle 0,2|\int_{-\infty}^\infty\frac{d\tilde{p}}{2\pi}\int_{-\infty}^\infty\frac{d\tilde{k}}{2\pi}\frac{e^{i(\tilde{k}+\tilde{p})(z_2-z_1)}}{(\alpha (\tilde{k}^2+\tilde{p}^2)-2\omega_0-\omega_{02})(\alpha \tilde{k}^2+\omega_0-\omega_{12})}\bigg(\frac{1}{\alpha\tilde{k}^2+\omega_0-\omega_{12}}+\nonumber \\
 \frac{1}{\alpha \tilde{p}^2+\omega_0-\omega_{12}}\bigg) +h.c.
 \end{align}
The dimensionless $\tilde{p}$ integral can be evaluated exactly. Doing so, we have
  \begin{align}
 H_{|2,0\rangle\leftrightarrow|0,2\rangle}=
 -2g^4|2,0\rangle\langle 0,2|\int_{-\infty}^\infty\frac{d\tilde{k}}{2\pi}e^{i\tilde{k}(z_2-z_1)}\bigg(\frac{e^{-\frac{|z_2-z_1|}{\sqrt{\alpha}}\sqrt{2\omega_0+\alpha \tilde{k}^2-\omega_{02}}}}{\sqrt{\alpha}\sqrt{2\omega_0+\alpha \tilde{k}^2-\omega_{02}}}\frac{1}{(\alpha \tilde{k}^2+\omega_0-\omega_{12})^2}-\nonumber\\
 \frac{1}{(\alpha \tilde{k}^2+\omega_0-\omega_{12})}\frac{1}{\omega_0+\omega_{12}-\omega_{02}+\alpha \tilde{k}^2}\bigg(\frac{e^{-\frac{|z_2-z_1|}{\sqrt{\alpha}}\sqrt{2\omega_0+\alpha \tilde{k}^2-\omega_{02}}}}{\sqrt{\alpha}\sqrt{2\omega_0+\alpha \tilde{k}^2-\omega_{02}}}-\frac{e^{-\frac{|z_2-z_1|}{\sqrt{\alpha}}\sqrt{\omega_0-\omega_{12}}}}{\sqrt{\alpha}\sqrt{\omega_0-\omega_{12}}}\bigg)\bigg)+h.c.
 \end{align}
Here, we have assumed that $2\omega_{0}>\omega_{02}$ and $\omega_0>\omega_{12}$. These conditions are satisfied in the regime where we experimentally observe the two-photon avoided crossing. We are interested in determining how the interaction decays as a function of distance. Unfortunately, we are unaware of how to analytically evaluate the first two terms (the third term can be evaluated exactly). However, the integrand decreases exponentially as a function of $k$ for the first two terms. Thus, it is reasonable to take the integrand to be a constant value (i.e. the integrand value at $k=0$) over a small window, $\delta k$, about $\tilde{k}=0$ and zero otherwise. This small window is taken to be the momentum value for which the argument in the exponential equals equals one, i.e. $\delta k=2\sqrt{\frac{1}{|z_2-z_1|^2}-\frac{2\omega_0-\omega_{02}}{\alpha}}$. Evaluating the remaining integral gives,
 \begin{align}
 H_{|2,0\rangle\leftrightarrow|0,2\rangle}\approx-2g^4\frac{|2,0\rangle\langle 0,2|}{(\omega_0-\omega_{12})}\frac{\delta k}{2\pi}\frac{e^{-|z_2-z_1|\sqrt{\frac{2\omega_0-\omega_{02}}{\alpha}}}}{\sqrt{\alpha}\sqrt{2\omega_0-\omega_{02}}}\bigg(\frac{1}{(\omega_0-\omega_{12})}-
\frac{1}{\omega_0+\omega_{12}-\omega_{02}}\bigg)\nonumber\\
+g^4|2,0\rangle\langle 0,2|\frac{1}{\sqrt{\alpha}(\omega_{02}-2\omega_{12})}\bigg(\frac{e^{-|z_2-z_1|\sqrt{\frac{\omega_0-\omega_{12}}{\alpha}}}}{\sqrt{\omega_0-\omega_{12}}}-\frac{e^{-|z_2-z_1|\sqrt{\frac{\omega_0+\omega_{12}-\omega_{02}}{\alpha}}}}{\sqrt{\omega_0+\omega_{12}-\omega_{02}}}\bigg)\frac{e^{-\frac{|z_2-z_1|}{\sqrt{\alpha}}\sqrt{\omega_0-\omega_{12}}}}{\sqrt{\alpha}\sqrt{\omega_0-\omega_{12}}}+h.c.
 \end{align}
We see that every term decays exponentially as a function of $|z_2-z_1|$, thus the effective interaction between the $|2,0\rangle$ and the $|0,2\rangle$ states decays exponentially as well.
 

\end{document}